\documentstyle[aps,version2,preprint]{revtex}    % preprint form
\begin{document}

\draft

\begin{title}
Topological ground-state excitations and symmetry\\ 
in the many-electron one-dimensional problem 
\end{title}

\author{J. M. P. Carmelo and N. M. R. Peres} 
\begin{instit}
Department of Physics, University 
of \'Evora, Apartado 94, P - 7001 \'Evora Codex, Portugal\\
and Institute for Scientific Interchange Foundation,
Villa Gualino, I - 10133 Torino, Italy 
\end{instit}
\receipt{31 May 1995}
% hooka
\begin{abstract}
We consider the Hubbard chain in a magnetic field and chemical 
potential. We introduce a pseudohole basis where all states are 
generated from a single reference vacuum. This allows the
evaluation for all sectors of Hamiltonian symmetry of the model 
of the expression of the $\sigma $ electron and hole operators at 
Fermi momentum $\pm k_{F\sigma }$ and vanishing excitation
energy in terms of pseudohole operators. 
In all sectors and to leading order in the excitation energy the
electron and hole are constituted by one $c$ pseudohole, one 
$s$ pseudohole, and one {\it topological momenton}. These
three quantum objects are confined in the electron or hole and
cannot be separated. We find that the set of 
different pseudohole types which in pairs constitute the two 
electrons and two holes associated with the transitions
from the $(N_{\uparrow },N_{\downarrow })$ ground state to the
$(N_{\uparrow }+1,N_{\downarrow })$, $(N_{\uparrow 
},N_{\downarrow }+1)$ and $(N_{\uparrow }-1,N_{\downarrow })$, 
$(N_{\uparrow },N_{\downarrow }-1)$ ground states, respectively, 
transform in the representation of the symmetry group of the 
Hamiltonian in the initial-ground-state sector of parameter space.
We also find the pseudohole generators for the half-filling holon 
and zero-magnetic-field spinon. The pseudohole basis introduced
in this paper is the only suitable for the extension of the
present type of operator description to the whole Hilbert space.
\end{abstract}
\renewcommand{\baselinestretch}{1.65546}   % for preprint style only

%\pacs{PACS numbers: 75.10 Lp, 72.15. Nj, 05.30. Fk, 03.65. Ca}

\narrowtext

%%%%%%%%%%%%%%%%%%%%%%%%%%%%%%%%%%%%%%%%%%%%%%%%%%%%%%%%%%%%%%%%
\section{INTRODUCTION}

In contrast to three-dimensional Fermi liquids \cite{Pines,Baym}, 
the low-energy excitations of one-dimensional many-electron 
quantum problems are at zero magnetic field characterized by a 
charge - spin separation \cite{Solyom,Voit,Metzner}. This can be 
interpretated in terms of holon and spinon modes \cite{Anderson}.
On the other hand, at finite magnetic fields the charge and
spin separation is replaced by a more exotic $c$ and
$s$ separation \cite{Carmelo94c}. Here $c$ and $s$ refer
to orthogonal small-momentum and low-energy modes which
couple both to the charge and spin channels. 
 
In the case of many-particle problems solvable by Bethe ansatz 
(BA) \cite{Bethe,Yang,Korepinrev} the low-energy spectrum can 
be studied explicitly. The BA solution of the Hubbard chain 
\cite{Lieb} at zero magnetic field and chemical potential has
allowed the identification and study of the holon and spinon 
excitations and corresponding symmetry transformations  
\cite{Essler}. (The spinon excitations of the Hubbard
chain are similar to the corresponding spinon excitations of 
the spin $1/2$-isotropic Heisenberg chain \cite{Faddeev}.)
However, the exact expression of the electron operator in terms of 
holon and spinon operators remains an open problem of some
complexity. Its solution requires an operator representation 
for the generators of the holon and spinon excitations. Even 
if such operator representation is obtained, a second problem is 
expressing these generators in terms of electronic operators. 
Moreover, the relation between Hamiltonian symmetry and the 
transformation of the elementary excitations has not been studied for 
finite magnetic fields or (and) densities away of half filling. In this 
case the holon and spinon picture breaks down, as we show in this 
article. 

For the general case of the Hubbard chain at finite magnetic field 
and chemical potential both the non-lowest-weight states (non-LWS's)
and non-highest weight states (non-HWS's) of the eta-spin and 
spin algebras \cite{Heilmann,Lieb89,Yang89,Zhang,Korepin,Ostlund}
have energy gaps relative to the corresponding canonical-ensemble
ground state \cite{Carmelo95}. The LWS's and HWS's of these algebras
can be classified into two types, the states I (or LWS's I and
HWS's I) and the states II (or LWS's II and HWS's II). While
the Hamiltonian eigenstates I are described only by real
BA rapidities, all or some of the rapidities associated
with the eigenstates II are complex and non real. Since
for finite magnetic field and chemical potential the
states II have energy gaps relative to the corresponding
canonical-ensemble ground state, in that case the low-energy physics
is exclusively determined by the states I \cite{Carmelo95}.

The pseudoparticle perturbation theory introduced in Refs. 
\cite{Carmelo90,Carmelo91,Carmelo92,Carmelo92b,Carmelo93}
and developed in a suitable operator basis in Refs.
\cite{Carmelo94c,Carmelo93b,Carmelo94,Carmelo94b} 
refers to the Hilbert subspace spanned by the Hamiltonian
eigenstates I. Rather than holons and spinons, at finite values
of the magnetic field and chemical potential and
at constant electronic numbers the low-energy excitations I  
are described by pseudoparticle-pseudohole processes relative to
the canonical-ensemble ground state. The latter state,
as well as all excited states I with the same
electron numbers, are simple Slater determinants of pseudoparticle
levels \cite{Carmelo94c,Carmelo95,Carmelo93b,Carmelo94,Carmelo94b}. 

One of the goals of this paper is introducing an alternative 
pseudohole representation which generates both the LWS's I
and the HWS's I from the same reference vacuum. Importantly,
it will be shown elsewhere that all Hamiltonian eigenstates
can be generated from the new pseudohole vacuum. In the case
of the states I the present pseudohole picture is alternative 
to the pseudoparticle description. However, for the extension
of our description to the whole Hilbert space the pseudohole 
picture is far more suitable. The description of 
the one-dimensional quantum problem in terms of 
forward-scattering-interacting pseudoparticles was introduced 
in Ref. \cite{Carmelo90} for contact-interaction soluble problems. 
Shortly after, the same kind of ideas were applied to 
$1/r^2$-interaction integrable models \cite{Haldane91,Shastry}. 
In the case of electronic models as the Hubbard chain such 
description is very similar to Fermi-liquid theory, except for two 
main differences: (i) the $\uparrow $ and $\downarrow $ 
quasiparticles are replaced by the $c$ and $s$ pseudoparticles 
and (ii) the discrete pseudoparticle momentum (pseudomomentum) is 
of the usual form $q_j={2\pi\over {N_a}}I_j^{\alpha}$ but the 
numbers $I_j^{\alpha}$ are not always integers. They are integers 
or half-integers depending on whether the number of $\uparrow$ or 
$\downarrow$ electrons in the system is even or odd. This plays a 
central role in the ground-state -- ground-state transitions 
\cite{Carmelo95b} we study in Sec. III.
The $c$ and $s$ pseudoparticles are non-interacting at the 
small-momentum and low-energy fixed point and the energy spectrum 
is described in terms of their bands. At higher 
energies and (or ) large momenta the pseudoparticles 
start to interact via zero-momentum transfer forward-scattering 
processes. As in a Fermi liquid, these are associated
with $f$ functions.

On the other hand, the transitions between states I
differing by odd electron numbers have a topological
character. This follows from the changes from
integers (or half integers) to half integers (or integers)
of the pseudoparticle quantum numbers $I_j^{\alpha}$.
Such topological excitations involve a global
shift of the corresponding pseudo-Fermi sea 
\cite{Carmelo94c,Carmelo95b} combined with a process of 
removing (adding) or adding (removing) pseudoparticles
(pseudoholes).

In general, all excitations involving states I can be decoupled 
into two types of transitions: (a) a topological ground-state -- 
ground-state transition which changes the pseudoparticle (and 
electron) numbers and involves pseudo-Fermi sea global shifts, 
which we call {\it topological momentons}; and (b) a pseudoparticle 
- pseudohole excitation relative to the final ground state.
In Ref. \cite{Carmelo95b} it was shown that the generators of 
the transitions (a) are $\sigma $ quasiparticles or quasiholes. 
Except for a vanishing renormalization factor, these entities are 
low-energy electrons or holes, respectively. The presence of such 
factor implies that in the limit of vanishing 
excitation energy there is a singular relation between these 
quasiparticles (quasiholes) and the electrons (holes). 
By expressing these quasiparticles or quasiholes in terms of 
pseudparticle operators one can then find the pseudoparticle 
contents of the electron or hole. 

In this paper we introduce a pseudohole description which allows 
the generalization of the results of Ref. \cite{Carmelo95b}
concerning the electron -- quasiparticle 
transformation to all sectors of symmetry of the
Hubbard chain in a magnetic field and chemical
potential \cite{Carmelo95c}. Only this description is 
suitable for the study of the interplay between Hamiltonian
symmetry and the transformation laws of the elementary
excitations which constitute the electrons and holes of
vanishing excitation energy. In contrast to the low-energy excitations
at constant electron numbers, which at zero magnetic
field and (or ) chemical potential can be states II
or non-LWS's and non-HWS's, all ground states are
states I \cite{Carmelo95}. This is also valid for canonical 
ensembles corresponding to zero magnetic field and (or )
chemical potential where the symmetry of the quantum problem
is higher. Therefore, all ground-state -- ground-state
transitions can be described in terms of $c$ and $s$ pseudoparticles
or pseudoholes. 

In the case of integrable models of simple Abelian $U(1)$
symmetry the elementary excitations are generated by a single 
type of pseudoparticles (pseudoholes ) \cite{Anto}. On the other hand,
in the present case of the Hubbard chain we have shown 
\cite{Carmelo95} that in each of the four sectors of Hamiltonian symmetry 
$U(1)\otimes U(1)$ there is one branch of $c$ pseudoparticles and 
one branch of $s$ pseudoparticles which describe the low-energy 
physics. In terms of pseudoholes the description of the
states I involves four branches of $\alpha ,\beta$ pseudoholes, 
where $\alpha =c,s$ and $\beta=\pm {1\over 2}$, as we discuss in 
future sections. In the present case of LWS's I and (or) HWS's I we 
have that $\beta=sgn (\eta_z){1\over 2}$ for $\alpha =c$ and 
$\beta=sgn (S_z){1\over 2}$ for $\alpha =s$. Therefore, 
in each $(l,l')=(sgn (\eta_z)1,sgn (S_z)1)$ sector of Hamiltonian 
symmetry $U(1)\otimes U(1)$ \cite{Carmelo95} only the $c,{l\over 
2}$ and $s,{l'\over 2}$ pseudohole branches are involved in the 
description of the corresponding states I. 

We express the low-energy $\sigma $ electron and hole of momentum 
$\pm k_{F\sigma }$ in terms of pseudoholes for the nine sectors of 
Hamiltonian symmetry of the quantum problem. In all sectors both the 
electron and the hole are constituted by one topological momenton, 
one $c$ pseudohole, and one $s$ pseudohole which cannot be 
disassociated and are confined within the electron or hole.
In the very particular limit of half filling and zero magnetic 
field we recover the holon and spinon picture and the associate 
symmetry transformation laws already found in Ref. \cite{Essler}.
In addition, we were able to find the operator 
generators of the holon and spinon excitations. The electron 
and hole contains one anti holon and holon, respectively, and
one spinon. For instance, we find that the half-filling 
$(\eta =1/2;S=0;\eta_z=-1/2;S_z=0)$ holon [and the 
$(\eta =1/2;S=0;\eta_z=1/2;S_z=0)$ anti holon] \cite{Essler} of 
lowest energy is constituted by one $c$ topological momenton and 
one $c,-{1\over 2}$ pseudohole [and one $c,{1\over 2}$ pseudohole]. The 
zero-magnetization $(\eta =0;S =1/2;\eta_z =0;S_z=-1/2)$ spinon 
of lowest energy [and the $(\eta= 0;S=1/2;\eta_z =0;S_z=1/2)$ 
spinon] \cite{Essler} is identified with one $s,{1\over 2}$ pseudohole 
[with one $s,-{1\over 2}$ pseudohole].

We also generalize the relation between the transformation
laws of the elementary excitations and the symmetry of
the Hamiltonian to all sectors of parameter space. 
We find that the set of different pseudohole types which
constitute the $\sigma $ electron and hole, ie which 
generate from the $(N_{\uparrow },N_{\downarrow })$ ground state the
$(N_{\uparrow }+1,N_{\downarrow })$, $(N_{\uparrow },N_{\downarrow }+1)$,
and $(N_{\uparrow }-1,N_{\downarrow })$, and $(N_{\uparrow 
},N_{\downarrow }-1)$ ground states, respectively, transform in 
the representation of the symmetry group of the Hamiltonian in 
the initial-ground-state sector of parameter space.
Our results also reveal that the occurence in the Hubbard chain
of $\eta $ pairing \cite{Heilmann,Yang89} at momentum $\pm\pi$  
is a necessary condition for ground states differing in the 
number of $\sigma $ electrons by one to have as relative momentum
the Fermi momentum $k_{F\sigma }$ or $-k_{F\sigma }$. 

In Section II we introduce the pseudohole description for 
the four sectors of parameter space of Hamiltonian symmetry 
$U(1)\otimes U(1)$ 
\cite{Carmelo94c,Carmelo95,Carmelo93b,Carmelo94,Carmelo94b}. 
For the case of ground states this 
description refers to all nine Hamiltonian symmetry sectors. 
We also introduce the pseudohole vacuum which we find to be the $SO(4)$ 
ground state. In the pseudohole representation all ground states and 
remaining states I are simple Slater determinant of pseudohole levels. 
We also construct the momentum operator and evaluate the momentum 
expression for all ground states of the problem. 

In Sec. III we express for all sectors of parameter space  
the momentum $\pm k_{F\sigma }$ electron and hole operators in terms 
of pseudohole operators and topological momenton operators. We 
also show that the usual holons and spinons are particular
limits of our pseudohole excitations and study the interplay 
between Hamiltonian symmetry and the set of pseudoholes 
which describes the $\sigma $ electrons and holes in each
canonical ensemble.

Finally, in Sec. IV we present the discussion and concluding 
remarks.

%%%%%%%%%%%%%%%%%%%%%%%%%%%%%%%%%%%%%%%%%%%%%%%%%%%%%%%%%%%%%%%%
\section{THE PSEUDOHOLE BASIS IN THE FOUR $U(1)\otimes U(1)$ 
         SECTORS}

We consider the Hubbard model \cite{Lieb,Carmelo92,Carmelo94,Frahm}
in one dimension with a finite chemical potential $\mu$ and in the
presence of a magnetic field $H$,

\begin{equation}
\hat{H} = \hat{H}_{SO(4)} + 2\mu\hat{\eta }_z + 
2\mu_0 H\hat{S}_z \, , 
\label{Hamilt}
\end{equation}
where

\begin{equation}
\hat{H}_{SO(4)} = -t\sum_{j,\sigma}
\left[c_{j\sigma}^{\dag }c_{j+1\sigma} +
c_{j+1\sigma}^{\dag }c_{j\sigma}\right] +
U\sum_{j} [c_{j\uparrow}^{\dag }c_{j\uparrow} - 1/2]
[c_{j\downarrow}^{\dag }c_{j\downarrow} - 1/2] \, ,
\label{Hamilt}
\end{equation}
and

\begin{equation}
\hat{\eta}_z = -{1\over 2}[N_a - \sum_{\sigma}\hat{N}_{\sigma }] \, ,
\hspace{1cm} 
\hat{S}_z = -{1\over 2}\sum_{\sigma}\sigma\hat{N}_{\sigma } \, ,
\end{equation}
are the diagonal generators of the $SU(2)$ eta-spin
and spin algebras, respectively \cite{Yang89,Korepin}. 

In equations $(1)-(3)$ the operator $c_{j\sigma}^{\dagger}$ 
and $c_{j\sigma}$ creates and annihilates, respectively, one 
electron of spin projection $\sigma$ ($\sigma$ refers to the spin 
projections $\sigma =\uparrow\, ,\downarrow$ when used as an 
operator or function index and is given by $\sigma =\pm 1$ 
otherwise) at the site $j$, $\hat{N}_{\sigma }=\sum_{j} 
c_{j\sigma }^{\dagger }c_{j\sigma }$ is the number operator for 
$\sigma$ spin-projection electrons, and $t$, $U$, $\mu$, $H$, and 
$\mu _0$ are the first-neighbor transfer integral, the onsite 
Coulomb interaction, the chemical potential, the magnetic field, 
and the Bohr magneton, respectively. 

There are $N_{\uparrow}$ up-spin electrons and $N_{\downarrow}$ 
down-spin electrons in the chain of $N_a$ sites and
with lattice constant $a$ associated with the model
$(1)$. We use periodic boundary conditions and
consider $N_a$ to be even and when $N=N_a$ (half filling)
both $N_{\uparrow}$ and $N_{\downarrow}$ to be odd
and employ units such that $a=t=\mu_0=\hbar =1$. 
When $N_{\sigma }$ is odd the Fermi momenta are given by 

\begin{equation}
k_{F\sigma }^{\pm }=\pm {\pi\over {N_a}}\left[N_{\sigma } 
-1\right] \, ,
\end{equation}
and when $N_{\sigma }$ is even the Fermi momenta are given by

\begin{equation}
k_{F\sigma}^{+} = {\pi\over {N_a}}N_{\sigma } \, ,
\hspace{1cm}
k_{F\sigma}^{-} = - {\pi\over {N_a}}\left[N_{\sigma } 
-2\right] \, ,
\end{equation}
or by

\begin{equation}
k_{F\sigma}^{+} = {\pi\over {N_a}}\left[N_{\sigma } 
-2\right] \, ,
\hspace{1cm}
k_{F\sigma}^{-} = - {\pi\over {N_a}}N_{\sigma } \, .
\end{equation}
Often we can ignore the ${1\over {N_a}}$ corrections
in the right-hand side (rhs) of Eqs. $(4)-(6)$ and consider
$k_{F\sigma}^{\pm}\simeq\pm k_{F\sigma}=\pm \pi n_{\sigma }$
and $k_F=[k_{F\uparrow}+k_{F\downarrow}]/2=\pi n/2$, where
$n_{\sigma}=N_{\sigma}/N_a$ and $n=N/N_a$. The
dimensionless onsite interaction is $u=U/4t$.

The two $SU(2)$ algebras -- eta spin 
and spin  -- have diagonal generators given by Eq. (3) and 
off-diagonal generators \cite{Essler,Korepin}

\begin{equation}
\hat{\eta}_{-} = \sum_{j} (-1)^j c_{j\uparrow}c_{j\downarrow}
\, , \hspace{1cm}   
\hat{\eta}_{+} = \sum_{j} (-1)^j c^{\dagger }_{j\downarrow}
c^{\dagger }_{j\uparrow} \, ,
\end{equation}
and

\begin{equation}
\hat{S}_{-} = \sum_{j} c^{\dagger }_{j\uparrow}c_{j\downarrow}
\, , \hspace{1cm}     
\hat{S}_{+} = \sum_{j} c^{\dagger }_{j\downarrow}
c_{j\uparrow} \, ,
\end{equation}
respectively. In the absence of the chemical-potential and 
magnetic-field terms the Hamiltonian $(1)$ reduces to $(2)$ and
has $SO(4) = SU(2) \otimes SU(2)/Z_2$ symmetry
\cite{Heilmann,Lieb89,Yang89,Zhang,Korepin,Ostlund}.
Since $N_a$ is even, the operator 
${\hat{\eta}}_z+{\hat{S}}_z$ [see Eq. $(3)$] has only integer 
eigenvalues and all half-odd integer representations of
$SU(2) \otimes SU(2)$ are projected out
\cite{Essler,Korepin}. 

For finite values of both the magnetic field and chemical
potential the symmetry of the quantum problem is reduced to 
$U(1)\otimes U(1)$, with $\hat{\eta}_z$ and $\hat{S}_z$ 
commuting with $\hat{H}$. The eigenvalues ${\eta}_z$ and $S_z$ 
determine the different symmetries of the Hamiltonian
$(1)$. When $\eta_z\neq 0$ and $S_z\neq 0$ the symmetry is 
$U(1)\otimes U(1)$, for $\eta_z= 0$ and $S_z\neq 0$ (and 
$\mu =0$) it is $SU(2)\otimes U(1)$, when $\eta_z\neq 0$ 
and $S_z= 0$ it is $U(1)\otimes SU(2)$, and at $\eta_z= 0$ 
and $S_z= 0$ (and $\mu =0$) the Hamiltonian symmetry is $SO(4)$. 

There are four $U(1)\otimes U(1)$ sectors 
of parameter space which correspond to $\eta_z< 0$ and $S_z< 0$; 
$\eta_z< 0$ and $S_z> 0$; $\eta_z> 0$ and $S_z< 0$; and 
$\eta_z> 0$ and $S_z> 0$. We follow Ref. \cite{Carmelo95}
and refer these sectors by $(l,l')$ where

\begin{equation}
l=sgn (\eta_z)1 \, ; \hspace{2cm} l'=sgn (S_z)1 \, .
\end{equation}
The sectors $(-1,-1)$; $(-1,1)$; $(1,-1)$; and $(1,1)$
refer to electronic densities and spin densities $0<n<1$ and  
$0<m<n$; $0<n<1$ and $-n<m<0$; $1<n<2$ and $0<m<(2-n)$;
and $1<n<2$ and $-(2-n)<m<0$, respectively.
  
There are two $(l')$ sectors of $SU(2)\otimes U(1)$ Hamiltonian 
symmetry [and two $(l)$ sectors of $U(1)\otimes SU(2)$ Hamiltonian 
symmetry] which correspond to $S_z< 0$ and $S_z> 0$ for
$l'=-1$ and $l'=1$, respectively, (and to $\eta_z< 0$ and $\eta_z> 0$ 
for $l=-1$ and $l=1$, respectively). There is one $SO(4)$ 
sector of parameter space [which is constituted only
by the $\eta_z= 0$ (and $\mu =0$) and $S_z= 0$ canonical
ensemble].

In Ref. \cite{Carmelo95} we have considered the BA solution for 
the model $(1)$ associated with each of the four sectors $(l,l')$
of Hamiltonian symmetry $U(1)\otimes U(1)$. As we have mentioned 
above, the pseudohole algebra introduced in the present paper is 
more suitable for the description of the non-LWS's and non-HWS's
eta-spin and spin multiplets which are not contained in the BA 
solution and whose study will be presented elsewhere
\cite{Carmelo95c}. Although for the case of the Hilbert subspace 
spanned by the states I both the pseudoparticle representation 
of Ref. \cite{Carmelo95} and the present pseudohole description 
are valid, we use here the latter which also simplifies the 
description of these states which in the pseudohole basis are 
generated from a single reference vacuum. (Instead of four 
pseudoparticle vacua \cite{Carmelo95}.)
 
In the Appendix A we discuss the connection of the present 
pseudohole description of the states I to the one of Ref. 
\cite{Carmelo95}. In that Appendix we also relate the BA equations to 
the pseudohole basis introduced in this section and present the 
Hamiltonian in that basis which includes the pseudohole 
dispersion relations and $f$ functions.

The pseudohole description we introduce below and in  
Appendix A includes four pseudohole branches which we denote in 
general by $\alpha ,\beta$ pseudoholes. Here $\alpha =c,s$ 
\cite{Carmelo94c,Carmelo93b,Carmelo94,Carmelo94b} 
and $\beta=\pm {1\over 2}$. The colors $c$ 
and $s$ and quantum numbers $\beta=\pm {1\over 2}$ which label the
four pseudohole branches also label the Hamiltonian eigenstates
I which correspond to different $\alpha ,\beta$ pseudohole
distribution occupancies. (About the relation between the
$\alpha ,\beta$ pseudoholes and the $(l,l')$ pseudoparticles 
of Ref. \cite{Carmelo95} see Appendix A.) 

Both in the case of the present states I and of the associate 
non-LWS's and non-HWS's \cite{Carmelo95c} the 
$\alpha $ orbitals have $N_{\alpha }^*$ 
available pseudomomentum values. In the latter states these 
pseudomomentum values can either be empty (this means occupation 
by one $\alpha $ pseudoparticle), single occupied by one $\alpha, 
{1\over 2}$ pseudohole, or single occupied by one $\alpha, -{1\over 2}$ 
pseudohole. This reveals that in the general case we should
consider $\alpha $ pseudoparticles and $\alpha, \beta$ pseudoholes
but {\it no} $\alpha, \beta$ pseudoparticles. Therefore, in order to 
simplify the future generalization of the present results to the 
whole Hilbert space, we use in this paper the suitable $\alpha $ 
pseudoparticle and $\alpha ,\beta$ pseudohole description. In contrast 
to the above non-LWS's and non-HWS's, in the case of the states I 
and $(l,l')$ sectors each of the $N_{\alpha }^*$ available
pseudomomentum values can either be empty (this means occupation by 
one $\alpha $ pseudoparticle) or single occupied by one $\alpha, 
\beta$ pseudohole, where $\beta$ is fixed and given by $\beta 
={l\over 2}$ for $\alpha =c$ or $\beta ={l'\over 2}$ for $\alpha =s$. 
The general expressions for $N_{\alpha }^*$ and number of $\alpha $
pseudoparticles, $N_{\alpha }$, are

\begin{equation}
N_c^* = N_a \, , \hspace{1cm} N_s^* = 
{1\over 2}\left[N_a - 2(\eta -S)\right] \, ,
\end{equation}
and

\begin{equation}
N_c = N_a - 2\eta \, , \hspace{1cm} N_s = 
{1\over 2}\left[N_a - 2(\eta +S)\right] \, ,
\end{equation}
respectively.

The results of Sec. III reveal that the existence of four types 
of $\alpha ,\beta$ pseudoholes with $\alpha =c,s$ and 
$\beta=\pm {1\over 2}$ is consistent with the Hamiltonian 
symmetry. Note that because the description of a state I of
the $(l,l')$ sector involves only two out of these four 
branches, namely the $c,\beta ={l\over 2}$ and $s,\beta =
{l'\over 2}$ pseudoholes, 
in the associate Hilbert subspace the pseudohole quantum
number $\beta $ is directly related to and determined by
the quantities of Eq. $(9)$. This is not however a general 
property. For instance, in the case of the non-LWS's and non-HWS's 
we will study in Ref. \cite{Carmelo95c} $\beta$ has no direct 
relation to the numbers of Eq. $(9)$. Also
the topological transitions of Sec. III connect states I
belonging to different sectors and involve three or four
different branches of $\alpha ,\beta$ pseudoholes.

Let us denote the pseudohole creation and annihilation operators by
$a^{\dag }_{q,\alpha ,\beta}$ and $a_{q,\alpha ,\beta}$, respectively. 
They obey the anticommutative algebra

\begin{equation}
\{a^{\dag }_{q,\alpha ,\beta},a_{q' ,\alpha' ,\beta'}\} 
= \delta_{q,q'}\delta_{\alpha ,\alpha'}\delta_{\beta,\beta'} \, , 
\end{equation}
and

\begin{equation}
\{a^{\dag }_{q,\alpha ,\beta},a^{\dag }_{q',\alpha',\beta'}\} 
= \{a_{q,\alpha ,\beta},a_{q',\alpha',\beta'}\} 
= 0 \, . 
\end{equation}

As we have mentioned in Sec. I, the discrete pseudomomentum 
values are 

\begin{equation}
q_j={2\pi\over 
{N_a}}I_j^{\alpha } \, ,
\end{equation}
where $I_j^{\alpha }$ are {\it consecutive} integers 
or half integers. There are $N_{\alpha }^*$ possible
$I_j^{\alpha }$ values, the number of $\alpha $ pseudomomentum
orbitals $N_{\alpha }^*$ being given in Eq. $(10)$.
From the point of view of the pseudoholes, a state I is specified by 
the distribution of $N_{\alpha }$ unoccupied values
over the $N_{\alpha }^*$ available values. These unoccupied
values correspond to the $N_{\alpha }$ $\alpha $ pseudoparticles, 
the number $N_{\alpha }$ depending on $\eta $ and $S$ and being
given in Eq. $(11)$. The numbers  $I_j^c$ are integers (or half 
integers) for $N_s$ even (or odd), and $I_j^s$ are integers (or 
half integers) for $N_s^*$ odd (or even).

There are $N_{\alpha }^h=N_{\alpha }^*-N_{\alpha }$ occupied 
values, which following Eqs. $(10)$ and $(11)$ are given by 

\begin{equation}
N_c^h = 2\eta \, , \hspace{1cm} N_s^h = 2S \, .
\end{equation}
In the general case including both the states I and 
the corresponding non-LWS's and non-HWS's \cite{Carmelo95c}
we have that

\begin{equation}
N_{\alpha }^h = \sum_{\beta=\pm {1\over 2}} N_{\alpha ,\beta}^h \, ,
\end{equation}
where $N_{\alpha ,\beta}^h$ is the number of $\alpha ,\beta$ 
pseudoholes. In the present case of the states I and $(l,l')$
sectors the summation of Eq. $(16)$ simplifies to
$N_{\alpha }^h = N_{\alpha ,\beta}^h$, with $\beta={l\over 2}$
for $\alpha=c$ and $\beta={l'\over 2}$ for $\alpha=s$.
This selection rule also simplifies operator expressions
including $\beta$ summations, as we discuss in Appendix A.

In the four $(l,l')$ sectors the Hamiltonian eigenstates I 
are simple Slater determinants of $c,\beta$ and $s,\beta$ 
pseudohole levels. As it becomes obvious from Eq. $(15)$, the 
pseudohole vacuum is the $SO(4)$ ground state, which is the only 
existing state I of the corresponding canonical ensemble. On the other
hand, ground states of canonical ensembles belonging
the two $(l')$ sectors of Hamiltonian symmetry $SU(2)\otimes U(1)$
and the two $(l)$ sectors of Hamiltonian symmetry $U(1)\otimes SU(2)$
are Slater determinants of only $s,{l'\over 2}$ and 
$c,{l\over 2}$ pseudohole levels, respectively, and have no 
$c,{l\over 2}$ and $s,{l'\over 2}$ pseudoholes, respectively. 

One of the advantages of the pseudohole 
representation for the states I (and its multiplets 
\cite{Carmelo95c}) is that while the pseudoparticle Slater 
determinants of Refs. \cite{Carmelo95,Carmelo94}
refer to a different pseudoparticle vacuum in each of the four $(l,l')$ 
sectors the corresponding pseudohole expressions involve a single 
and unique vacuum which is common to all sectors. This is
the $SO(4)$ ground state.

Since the electron and hole studies of Sec. III refer to
ground-state -- ground-state transitions, in this section 
we focus our attention on ground states. In terms of pseudoholes 
the general ground-state expression studied in Ref. \cite{Carmelo95} 
and associated with the $(l,l')$ sector of Hamiltonian symmetry 
$U(1)\otimes U(1)$ reads 

\begin{equation}
|0;\eta_z,S_z\rangle = 
\prod_{q=q_{c }^{(-)}}^{{\bar{q}}_{Fc }^{(-)}} 
\prod_{q={\bar{q}}_{Fc }^{(+)}}^{q_{c }^{(+)}} 
a^{\dag }_{q,c,{l\over 2}} 
\prod_{q=q_{s}^{(-)}}^{{\bar{q}}_{Fs}^{(-)}} 
\prod_{q={\bar{q}}_{Fs}^{(+)}}^{q_{s}^{(+)}} 
a^{\dag }_{q,s,{l'\over 2}} 
|0;0,0\rangle \, ,
\end{equation}
where $l$ and $l'$ are given in Eq. $(9)$ and
$|0;0,0\rangle $ is the $SO(4)$ $\mu =0$ and
$H=0$ ground state. The pseudo-Fermi points $q_{F\alpha }^{(\pm)}$ 
and corresponding limits of the pseudo-Brillouin zones 
$q_{\alpha }^{(\pm)}$ are defined below. It is useful to introduce the
pseudohole-Fermi points ${\bar{q}}_{F\alpha }^{(\pm)}$
such that

\begin{equation}
{\bar{q}}_{F\alpha }^{(\pm)} = q_{F\alpha }^{(\pm)}
\pm {2\pi\over {N_a}} \, .
\end{equation}
When $N_{\alpha }$ is odd (or even) and $I_j^{\alpha }$ 
are integers (or half integers) the pseudohole-Fermi points are 
symmetric and given by

\begin{equation}
{\bar{q}}_{F\alpha }^{(+)}=-{\bar{q}}_{F\alpha }^{(-)} =
{\pi\over {N_a}}[N_{\alpha }+1] \, .
\end{equation}
On the other hand, when $N_{\alpha }$ is odd (or even) 
and $I_j^{\alpha }$ are half integers (or integers)
we have that either

\begin{equation}
{\bar{q}}_{F\alpha }^{(+)} = {\pi\over {N_a}}[N_{\alpha }
+ 2] \, , \hspace{1cm} 
{\bar{q}}_{F\alpha }^{(-)} = - {\pi\over {N_a}}N_{\alpha } \, ,
\end{equation}
or 

\begin{equation}
{\bar{q}}_{F\alpha }^{(+)} = {\pi\over {N_a}}N_{\alpha } 
\, , \hspace{1cm} 
{\bar{q}}_{F\alpha }^{(-)} = - {\pi\over {N_a}}[N_{\alpha } 
+ 2] \, . 
\end{equation}
The pseudo-Fermi points are defined by combining Eq. $(18)$
with Eqs. $(19)-(21)$.

The limits of the pseudo-Brillouin zones $q_{\alpha }^{(\pm)}$
involve the number of $\alpha $-pseudomomentum orbitals 
$N_{\alpha }^*$. When $N_{\alpha }^*$ is odd (or even) and 
$I_j^{\alpha }$ are integers (or half integers) the limits of 
the pseudo-Brillouin zones are symmetric and given by

\begin{equation}
q_{\alpha }^{(+)}=-q_{\alpha }^{(-)} =
{\pi\over {N_a}}[N_{\alpha }^*-1] \, .
\end{equation}
On the other hand, when $N_{\alpha }^*$ is odd (or even) 
and $I_j^{\alpha }$ are half integers (or integers)
we have either that

\begin{equation}
q_{\alpha }^{(+)} = {\pi\over {N_a}}N_{\alpha }^*
\, , \hspace{1cm} 
q_{\alpha }^{(-)} = - {\pi\over {N_a}}[N_{\alpha }^*-2] \, ,
\end{equation}
or 

\begin{equation}
q_{\alpha }^{(+)} = {\pi\over {N_a}}[N_{\alpha }^*-2] 
\, , \hspace{1cm} 
q_{\alpha }^{(-)} = - {\pi\over {N_a}}N_{\alpha }^* \, . 
\end{equation}

For the topological excitations studied in Sec. III the terms of
order ${1\over {N_a}}$ of the rhs of Eqs. $(19)-(24)$ play an important 
role. However, for many quantities these corrections are in the 
thermodynamic limit unimportant and we can consider instead

\begin{equation}
q_{F\alpha } = {\pi N_{\alpha }\over {N_a}}
\simeq \pm {\bar{q}}_{F\alpha }^{(\pm)}
\simeq \pm q_{F\alpha }^{(\pm)} \, ,
\end{equation}
and

\begin{equation}
q_{\alpha } = {\pi N_{\alpha }^*\over {N_a}}\nonumber
\simeq \pm q_{\alpha }^{(\pm)} \, .
\end{equation}

In all sectors of Hamiltonian symmetry there are states I. In the
particular case of the $SO(4)$ zero-chemical potential
and zero-magnetic field canonical ensemble there is only one state 
I. The study of the spectrum for the states II and non-LWS's
and non-HWS's multiplets reveals that this state I is nothing but 
the $SO(4)$ ground state \cite{Carmelo96}. The same applies to the sectors 
of Hamiltonian symmetry $SU(2)\otimes U(1)$ and $U(1)\otimes SU(2)$, 
the ground state being always a state I. (In addition, in these
sectors there is a large number of excited states I.)
While the description of the states II requires adding  
new ``heavy'' pseudoparticles onto the universal $SO(4)$
pseudohole vacuum \cite{Carmelo96}, all states I can be generated 
from that vacuum by distribution occupancies of $\alpha ,\beta$ 
pseudoholes only. In the case of the two $(l')$ $SU(2)\otimes U(1)$ 
sectors the ground state is both a LWS and HWS of the eta-spin algebra. 
Therefore, it is empty of $c$ pseudoholes and reads

\begin{equation}
|0;0,S_z\rangle = 
\prod_{q=q_{s}^{(-)}}^{{\bar{q}}_{Fs}^{(-)}} 
\prod_{q={\bar{q}}_{Fs}^{(+)}}^{q_{s}^{(+)}} 
a^{\dag }_{q,s,{l'\over 2}} 
|0;0,0\rangle \, .
\end{equation}

In the case of the $(l)$ $U(1)\otimes SU(2)$ sector 
the ground state is both a LWS and a HWS of the spin algebra
and is empty of $s$ pseudoholes. It reads

\begin{equation}
|0;\eta_z,0\rangle = 
\prod_{q=q_{c}^{(-)}}^{{\bar{q}}_{Fc}^{(-)}} 
\prod_{q={\bar{q}}_{Fc}^{(+)}}^{q_{c}^{(+)}} 
a^{\dag }_{q,c,{l\over 2}} |0;0,0\rangle \, .
\end{equation}

Finally, the $\eta =\eta_z =0$ (and $\mu =0$) and $S=S_z=0$ $SO(4)$ 
ground state is, at the same time, a LWS and HWS of both the eta-spin 
and spin algebras, ie following the notation of Ref. \cite{Carmelo95} 
it is a [LWS,LWS], a [LWS,HWS], a [HWS,LWS], and a 
[HWS,HWS]. Therefore, it is empty of both $c$ and $s$
pseudoholes and is the vacuum of the pseudohole theory.
All the remaining states I can be described by Slater determinants
of $\alpha ,\beta$ pseudoholes, as shown in Refs. \cite{Carmelo95,Carmelo94}
in terms of pseudoparticles.

We close this section by constructing the momentum
operator for the states I of all sectors of parameter space.
(This question was not addressed in Ref. \cite{Carmelo95}.) 
The $\pm\pi$ $\eta $-pairing \cite{Heilmann,Lieb89,Yang89} 
determines the value for the relative momentum of 
corresponding eta-spin LWSs and HWSs pairs of the same multiplet
family (and thus having the same value of $\eta $). In the two 
sectors $(-1,\pm 1)$ we have that $0<n<1$ and the momentum 
operator has the usual form \cite{Carmelo94}

\begin{equation}
\hat{P} = 
\sum_{q}q\left[1 - a^{\dag }_{q,c,-{1\over 2}}
a_{q,c,-{1\over 2}}\right] +
\sum_{q}q\left[1 - a^{\dag }_{q,s,\pm {1\over 2}}
a_{q,s,\pm {1\over 2}} \right] \, . 
\end{equation}

For each LWS I of the eta-spin algebra associated with the 
$(-1,l')$ sector there is one and only one HWS I of the eta-spin 
algebra in the corresponding $(1,l')$ sector which belongs
to the same tower. That HWS I is generated by acting onto the 
corresponding LWS I a suitable number of times the operator 
$\hat{\eta }_{+}$ of Eq. $(7)$. This operator has momentum
$\pm\pi$, ie when it acts once onto a state it generates a new
state with momentum $\pm\pi$ relatively to the initial state.

The fact that the $\eta $-pairing operators $(7)$ have in the
present model momentum $\pm\pi$ implies that the relative
momentum of the above LWS and HWS is either (a) a reciprocal-lattice
momentum, $G=2\pi j$ with $j=0,\pm 1,\pm 2,...$, or (b) a
reciprocal lattice momentum plus $\pm\pi$. The occurence 
of the cases (a) or (b) depends simply on the parity of the number of 
times needed to act the operator $\hat{\eta }_{+}$ onto the 
LWS to obtain the corresponding HWS: the cases (a) and (b) 
correspond to an even and odd number of times, respectively. It is
staightforward to find that when the total electron number $N$
associated with the canonical ensemble of the initial LWS
(which has the same parity as the electron number
of the final HWS) is either even or odd we have the case (a)
or (b), respectively. Since we choose the momentum of the final
HWS to belong the first Brillouin zone, this leads to the
following momentum-operator expressions for the $(1,\pm 1)$ 
sectors  

\begin{equation}
\hat{P} = 
\sum_{q}q\left[1 - a^{\dag }_{q,c,{1\over 2}}
a_{q,c,{1\over 2}}\right] +
\sum_{q}q\left[1 - a^{\dag }_{q,s,\pm {1\over 2}}
a_{q,s,\pm {1\over 2}} \right] 
\, , \hspace{1cm} N \hspace{0.25cm} even \, ,  
\end{equation}
and

\begin{equation}
\hat{P} = 
\pm\pi + \sum_{q}q\left[1 - a^{\dag }_{q,c,{1\over 2}}
a_{q,c,{1\over 2}}\right] +
\sum_{q}q\left[1 - a^{\dag }_{q,s,\pm {1\over 2}}
a_{q,s,\pm {1\over 2}}\right] \, , 
\hspace{1cm} N \hspace{0.25cm} odd \, , 
\end{equation}
where in equation $(31)$ we choose $\pi$ or $-\pi$ depending on which 
of these values provides the momentum $P$ of a given state in the 
first Brillouin zone. The momentum expressions $(29)-(31)$ are valid 
for the Hilbert subspace spanned by the states I. Since all ground 
states of canonical ensembles belonging the nine sectors of Hamiltonian 
symmetry of the present model are states I, the operator expressions 
$(29)-(31)$ provide the momenta of all ground states. These are always 
of the form $(29)$, $(30)$, or $(31)$. 

According to the integer or half-integer character of the 
$I_j^{\alpha }$ numbers we have four ``topological'' types of Hilbert 
subspaces I and corresponding ground states. Since that character depends 
on the parities of the $\eta $- and $S$-dependent
numbers $N_s^*$ and $N_s$ of Eqs. $(10)$ and $(11)$, we refer these 
subspaces by the parities of $N_s^*$ and $N_s$, respectively,
as: (A) even, even; (B) even, odd; (C) odd, even; and (D) odd, odd.
The ground-state momentum expression is different for each type of 
Hilbert sub space (A)-(D). While the ground states (A)-(C) are bi-degenerate,
the ground states corresponding to the Hilbert sub space (D)
have zero momentum and are non-degenerate. In Table I we
present the ground-state momentum values for the Hilbert 
sub spaces (A)-(D) in the four $(l,l')$ sectors. 

The $SU(2)\otimes U(1)$ and $U(1)\otimes SU(2)$ ground states $(27)$ 
and $(28)$, respectively, can belong to the Hilbert subspaces (A) and 
(C) only. These have momenta given in Table I [see (A) and (C)]. 
In the $SU(2)\otimes U(1)$ case we should use $2k_F=2\pi -2k_F=\pi$
in the momentum expressions of that Table. Finally, the $SO(4)$
ground state refers to the Hilbert subspace (C). Its has
zero momentum and is non degenerate.

As equations $(10)-(11)$ reveal, the eta spin
$\eta$, spin $S$ and the operators $\hat{\eta}_z$ and $\hat{S}_z$ 
have simple expressions in the pseudohole basis. The following 
expressions are both valid for the states I and for the non-LWS's 
and non-HWS's multiplets of these states \cite{Carmelo95c}:

\begin{equation}
\eta = {1\over 2}\sum_{\beta}N^h_{c,\beta} \, ,
\hspace{1cm} 
S = {1\over 2}\sum_{\beta}N^h_{s,\beta} \, .
\end{equation}
and

\begin{equation}
\hat{\eta}_z = \sum_{\beta}\beta\hat{N}^h_{c,\beta} \, ,
\hspace{1cm} 
\hat{S}_z = \sum_{\beta}\beta\hat{N}^h_{s,\beta} \, .
\end{equation}

In the Hilbert sub space spanned by the states I 
and at energies smaller than the gaps for the non-LWS's and non-HWS's 
and for the states II the Hubbard model $(1)$ can be written in 
the $\alpha ,\beta$ pseudohole basis. This expression
is presented in Appendix A. Its normal-ordered expression 
with respect to the suitable ground state $(17)$, Eqs. 
(A14)-(A15), has an infinite number of terms which correspond 
to increasing scattering orders. The perturbative character of 
the pseudohole basis \cite{Carmelo94c,Carmelo94,Carmelo94b} implies
that at low energies only the two terms of lower pseudohole 
scattering order are relevant. For a detailed study of the present 
quantum problem in the Hilbert subspace spanned by the states I see 
Refs. \cite{Carmelo91,Carmelo92,Carmelo92b,Carmelo94,Carmelo94b}
which consider the usual $(-1,-1)$ sector of Hamiltonian 
symmetry $U(1)\otimes U(1)$.

%%%%%%%%%%%%%%%%%%%%%%%%%%%%%%%%%%%%%%%%%%%%%%%%%%%%%%%%%%%%%%%%
\section{GROUND-STATE TRANSITIONS AND PSEUDOHOLE SYMMETRIES} 

The study of the interplay between the Hamiltonian symmetry and 
the transformation laws of the elementary excitations reveals
that symmetry is closely related to the low-energy electron
and hole {\it pseudohole} content. The main aim of this section 
is thus to relate the Hamiltonian symmetry to the transformations 
of the set of pseudoholes which form the electrons and holes 
in each parameter-space sector. This requires the generalization 
to all sectors of parameter space of the present quantum problem 
of recent results \cite{Carmelo95b} for the expression of the 
low-energy electron at the Fermi momentum $\pm k_{F\sigma }$ 
[$k_{F\sigma }^{\pm}$ if we use the discrete definition of Eqs. 
$(4),(6)$] in terms of pseudoholes. 

The studies of Ref. \cite{Carmelo95b} referred to the 
$(-1,-1)$ sector and expressed the electron and quasiparticle 
in terms of pseudoparticles. However, only the introduction of 
our pseudohole basis allows the study of the relation of the 
results of Ref. \cite{Carmelo95b} to Hamiltonian symmetry. 
In this section we express the $\sigma $ 
electrons and holes in terms of pseudoholes
for all nine sectors of parameter space. 
In the very particular limit of half filling and 
zero magnetization we recover the holon
and spinon symmetry results of Ref. \cite{Essler}. Moreover,
in that $SO(4)$ canonical ensemble our general operator study 
solves the problem of expressing the electron (the hole)
in terms of anti holons and spinons (of holons and spinons). We
find that the holons, anti holons, and spinons are closely related 
to limiting cases of our general pseudoholes.

To leading order in the excitation energy $\omega $ the 
$\sigma $ electron operator of momentum $\pm k_{F\sigma }$ is 
the product of a $\sigma $ quasiparticle operator of momentum 
$\pm k_{F\sigma }$ and a vanishing renormalization factor
\cite{Carmelo95b}. In spite of the singular character of this
electron -- quasiparticle transformation, which justifies the 
perturbative nature of the quantum problem in the pseudohole 
basis, the renormalization factor is
absorbed by the transformation. Therefore, expressing the 
quasiparticle in terms of pseudohole operators provides relevant 
physical information. This reveals that in terms of 
pseudoholes one electron is a topological excitation constituted
by one $c$ pseudohole, one $s$ pseudohole, and one large-momentum 
many-pseudohole topological excitation, the topological 
momenton. These three quantum objects are confined in the 
electron and cannot be separated.  

Given a ground state with electron numbers $(N_{\sigma},
N_{-\sigma })$, we find in this section that the set of all 
pseudoholes of different type which constitute, in pairs, 
the $\uparrow $ and $\downarrow $ electrons and $\uparrow $ and 
$\downarrow $ holes associated with the ground-state transitions  
$(N_{\uparrow},N_{\downarrow})\rightarrow 
(N_{\uparrow}\pm 1,N_{\downarrow})$ and $(N_{\uparrow},
N_{\downarrow})\rightarrow (N_{\uparrow},N_{\downarrow}\pm 1)$ 
transform as the symmetry group of the Hamiltonian $(1)$
in the corresponding sector of parameter space.

The study of the ground-state momentum expressions of the 
previous section reveals that the relative momentum of ground 
states differing in the number of $\sigma $ electrons by one equals 
the $U=0$ Fermi points, ie $\Delta P=\pm k_{F\sigma }$. We define 
the quasiparticle operator, ${\tilde{c}}^{\dag }_{k_{F\sigma },\sigma }$, 
which creates one quasiparticle with spin projection $\sigma$ and 
momentum $k_{F\sigma}$ as \cite{Carmelo95b} 

\begin{equation}
{\tilde{c}}^{\dag }_{k_{F\sigma},\sigma}
|0; N_{\sigma}, N_{-\sigma}\rangle =
|0; N_{\sigma} + 1, N_{\sigma}\rangle \, .
\end{equation}
The quasiparticle operator defines a one-to-one correspondence
between the addition of one electron to the system and the creation
of one quasiparticle. Exactly as is expected from the Landau theory
in three dimensions, in Appendix B we follow the $(-1,-1)$- sector 
study of Ref. \cite{Carmelo95b} and explain how the electronic 
excitation $c^{\dag }_{k_{F\sigma},\sigma}|0; N_{\sigma}, 
N_{-\sigma}\rangle$, defined at the Fermi momentum 
and small excitation energy $\omega $, contains a single 
quasiparticle. (We measure the energy $\omega $ from the 
initial-ground-state chemical potential.) In addition to 
the electron -- quasiparticle transformation (B1), we
also consider in that Appendix the hole -- quasihole 
transformation [see Eq. (B4)]. In spite of the singular
character of these transformations, the discussion of
Appendix B reveals that one quasiparticle (quasihole)
is basically one electron (hole). Therefore, we call
often below the quasiparticle (quasihole) as electron
(hole).

Let us then study the expression of the $\sigma $ quasiparticle and 
quasihole operators in the pseudohole basis for all sectors of 
Hamiltonian symmetry. Since we are discussing the problem of 
addition or removal of one particle the boundary conditions 
play a crucial role \cite{Carmelo94c,Carmelo95b}.
When we add or remove one electron from the many-body system we
have to consider the transitions between states with integer and
half-integer quantum numbers $I_j^{\alpha }$. The transition between
two ground states differing in the number of electrons
by one is then associated with two different processes: 
a backflow in the Hilbert space of the $\alpha ,\beta$ 
pseudoholes with a shift of all the pseudomomenta 
by $\pm\frac{\pi}{N_a}$ and the creation and (or) annihilation 
of one pair of $c$ and $s$ pseudoholes at the pseudo-Fermi points (or
at the limit of the pseudo-Brillouin zone for the $s$ pseudohole). 

The backflow associated with a shift of all the pseudomomenta
momenta by $\pm\frac{\pi}{N_a}$ is described by a topological
unitary operator such that 

\begin{equation}
V^{\pm}_{\alpha }a^{\dag }_{q,\alpha ,\beta}
V^{\mp}_{\alpha }= 
a^{\dag }_{q\mp {\pi\over {N_a}},\alpha ,\beta} \, .
\end{equation}
Obviously, the pseudohole vacuum is invariant under
this operator, ie 

\begin{equation}
V^{\pm}_{\alpha }|0;0,0\rangle =|0;0,0\rangle \, . 
\end{equation}
Using the same method as Ref. \cite{Carmelo95b}, we
find

\begin{equation}
V^{\pm}_{\alpha }=V_{\alpha } 
\left(\mp\frac{\pi}{N_a}\right) \, ,
\end{equation}
where

\begin{equation}
V_{\alpha }(\delta q) = \exp\left\{i\delta q 
G^h_{\alpha }\right\} \, ,
\end{equation}
and

\begin{equation}
G^h_{\alpha } = -i\sum_{q,\beta}\left[{\partial\over {\partial q}} 
a^{\dag }_{q,\alpha ,\beta}\right]a_{q,\alpha ,\beta} \, ,
\end{equation}
is the Hermitian generator of the $\pm\frac{\pi}{N_a}$
topological pseudomomentum translation. Adding all pseudohole 
contributions gives a large momentum. 
This large-momentum excitation induced by the operator
$(37)$ is the $\alpha $ topological momenton. That operator 
has the following discrete representation  

\begin{equation}
V^{\pm}_{\alpha } = \exp\left\{
-\sum_{q,\beta}a^{\dag }_{q\pm\frac{\pi}{N_a}
,\alpha ,\beta}a_{q,\alpha ,\beta}\right\} \, .
\end{equation}
Note that in the present case of the Hilbert subspace spanned by
the states I of the sector $(l,l')$ only the value 
$\beta={l\over 2}$ for $\alpha =c$ and the value
$\beta={l'\over 2}$ for $\alpha =s$ contributes to the 
$\beta$ summation of Eqs. $(39)$ and $(40)$, as discussed
in Sec. II and Appendix A.

In addition to the topological momenton, the quasiparticle
or quasihole excitation includes creation and (or ) 
annihilation of pseudoholes. The changes in the 
pseudohole and pseudoparticle numbers and the corresponding
changes in the values of $\eta$, $\eta_z$, $S$, and $S_z$ are 
given in Tables II and III for the ground-state -- ground-state 
transitions $(N_{\uparrow},N_{\downarrow})\rightarrow
(N_{\uparrow}\pm 1,N_{\downarrow})$ and
$(N_{\uparrow},N_{\downarrow})\rightarrow
(N_{\uparrow},N_{\downarrow}\pm 1)$, respectively.
 
We consider below the expressions for the quasiparticles 
${\tilde{c}}^{\dag }_{k_{F\uparrow},\uparrow}$ and
${\tilde{c}}^{\dag }_{k_{F\downarrow},\downarrow}$
associated with the transitions $(N_{\uparrow},
N_{\downarrow})\rightarrow (N_{\uparrow}+1,N_{\downarrow})$ and
$(N_{\uparrow},N_{\downarrow})\rightarrow
(N_{\uparrow},N_{\downarrow}+1)$, respectively, and the 
quasiholes ${\tilde{c}}_{k_{F\uparrow},\uparrow}$ and
${\tilde{c}}_{k_{F\downarrow},\downarrow}$
associated with the transitions
$(N_{\uparrow},N_{\downarrow})\rightarrow
(N_{\uparrow}-1,N_{\downarrow})$ and
$(N_{\uparrow},N_{\downarrow})\rightarrow
(N_{\uparrow},N_{\downarrow}-1)$, respectively. 

We emphasize that because the initial ground state for the above 
two quasiparticles and two quasiholes is the same, the $\sigma $ 
quasiparticle and $\sigma $ quasihole momenta differ by $\pm 
{2\pi\over {N_a}}$ [see Eqs. $(4)-(6)$]. Therefore, the corresponding 
quasiparticle and quasihole expressions are not related by 
an adjunt transformation. On the other hand, the operators 
$\tilde{c}^{\dag }_{\pm k_{F\sigma },\sigma }$ and
$\tilde{c}_{\pm k_{F\sigma },\sigma }$ associated with the 
transitions $(N_{\sigma},N_{-\sigma})\rightarrow
(N_{\sigma}+1,N_{-\sigma})$ and $(N_{\sigma}+1,N_{-\sigma})\rightarrow
(N_{\sigma},N_{-\sigma})$ are obviously related by
such transformation. In this case the initial (final)
ground state of the electrons (holes) is the final
(initial) ground state of the holes (electrons).  
Moreover, let us consider the set of four operators
$\tilde{c}^{\dag }_{\pm k_{F\uparrow},\uparrow }$,
$\tilde{c}_{\pm k_{F\uparrow },\uparrow }$, 
$\tilde{c}^{\dag }_{\pm k_{F\downarrow},\downarrow }$, and
$\tilde{c}_{\pm k_{F\downarrow },\downarrow }$
such that the creation operators act on the same
initial ground state $(N_{\uparrow},N_{\downarrow})$
transforming it in the ground states  $(N_{\uparrow }+1,
N_{\downarrow})$ and $(N_{\uparrow},N_{\downarrow} +1)$,
respectively, and the hole operators act on the corresponding
latter states giving rise to the original ground state.
Let us consider the reduced Hilbert subspace spanned
by these three ground states, the Fermi-point
discrete definitons $(4)-(6)$, and the pseudo-Fermi points
and pseudo-Brillouin-zone limits expressions $(18)-(24)$.
If we combine that with the electron and hole expressions 
introduced below, it is easy to show 
that the corresponding quasiparticle and quasihole operators 
$\tilde{c}^{\dag }_{\pm k_{F\uparrow},\uparrow }$,
$\tilde{c}_{\pm k_{F\uparrow },\uparrow }$, 
$\tilde{c}^{\dag }_{\pm k_{F\downarrow},\downarrow }$, and
$\tilde{c}_{\pm k_{F\downarrow },\downarrow }$ obey the usual
anticommutation relations.

Generalization of the results of Ref. \cite{Carmelo95b} leads 
to quasiparticle and quasihole operator expressions for all 
sectors. The two electrons and two holes refer to the same
initial ground state. The pseudo-Fermi points and pseudohole-Fermi 
points of the expressions below refer to that initial ground state.
On the other hand, $s$ pseudohole creation and annihilation
operators at the limits of the pseudo-Brillouin zones refer to the
final and initial ground states, respectively. In the case of
the $(l,l')$ sectors of Hamiltonian symmetry $U(1)\otimes U(1)$
we consider that the initial and final ground states belong
the same sector of parameter space. In the case of the $(l')$ 
sectors of Hamiltonian symmetry $SU(2)\otimes U(1)$ [or $(l)$ 
sectors of Hamiltonian symmetry $U(1)\otimes SU(2)$] we consider that 
the initial and final ground states belong to sectors of parameter
space characterized by the same value of $(l')$ [or $(l)$]. We 
present below the electron and hole expressions found for 
different initial ground states in the nine sectors of parameter 
space.

For initial ground states in the $(-1,-1)$ sector of Hamiltonian
symmetry $U(1)\otimes U(1)$ we find

\begin{equation}
\tilde{c}^{\dag }_{\pm k_{F\uparrow },\uparrow } =
a_{{\bar{q}}_{Fc}^{(\pm)},c,-{1\over 2}}V^{\pm}_{s}
a^{\dag }_{q_{s}^{(\pm)},s,-{1\over 2}} \, ,\hspace{1cm}
\tilde{c}^{\dag }_{\pm k_{F\downarrow },\downarrow } =
V^{\pm}_{c}a_{{\bar{q}}_{Fc}^{(\pm)},c.-{1\over 2}}
a_{{\bar{q}}_{Fs}^{(\pm)},s,-{1\over 2}} \, ,
\end{equation}
for the electrons and 

\begin{equation}
\tilde{c}_{\pm k_{F\uparrow },\uparrow } =
a^{\dag }_{q_{Fc}^{(\pm)},c,-{1\over 2}}V^{\mp}_{s}
a_{q_{s}^{(\pm)},s,-{1\over 2}} \, ,\hspace{1cm}
\tilde{c}_{\pm k_{F\downarrow },\downarrow } =
V^{\mp}_{c}a^{\dag }_{q_{Fc}^{(\pm)},c,-{1\over 2}}
a^{\dag }_{q_{Fs}^{(\pm)},s,-{1\over 2}} \, ,
\end{equation}
for the holes. 

For the $(-1,1)$ sector we find

\begin{equation}
\tilde{c}^{\dag }_{\pm k_{F\uparrow },\uparrow } =
V^{\pm}_{c}a_{{\bar{q}}_{Fc}^{(\pm)},c,-{1\over 2}}
a_{{\bar{q}}_{Fs}^{(\pm)},s,{1\over 2}} \, ,\hspace{1cm}
\tilde{c}^{\dag }_{\pm k_{F\downarrow },\downarrow } =
a_{{\bar{q}}_{Fc}^{(\pm)},c,-{1\over 2}}V^{\pm}_{s}
a^{\dag }_{q_{s}^{(\pm)},s,{1\over 2}} \, ,
\end{equation}
for the electrons and

\begin{equation}
\tilde{c}_{\pm k_{F\uparrow },\uparrow } =
V^{\mp}_{c}a^{\dag }_{q_{Fc}^{(\pm)},c,-{1\over 2}}
a^{\dag }_{q_{Fs}^{(\pm)},s,{1\over 2}} \, ,\hspace{1cm}
\tilde{c}_{\pm k_{F\downarrow },\downarrow } =
a^{\dag }_{q_{Fc}^{(\pm)},c,-{1\over 2}}V^{\mp}_{s}
a_{q_{s}^{(\pm)},s,{1\over 2}} \, ,
\end{equation}
for the holes.

In the $(1,-1)$ sector the result is

\begin{equation}
\tilde{c}^{\dag }_{\pm k_{F\uparrow },\uparrow } =
V^{\mp}_{c}a^{\dag }_{q_{Fc}^{(\pm)},c,{1\over 2}}
a^{\dag }_{q_{Fs}^{(\pm)},s,-{1\over 2}} \, ,\hspace{1cm}
\tilde{c}^{\dag }_{\pm k_{F\downarrow },\downarrow } =
a^{\dag }_{q_{Fc}^{(\pm)},c,{1\over 2}}V^{\mp}_{s}
a_{q_{s}^{(\pm)},s,-{1\over 2}} \, ,
\end{equation}
for the electrons and

\begin{equation}
\tilde{c}_{\pm k_{F\uparrow },\uparrow } =
V^{\pm}_{c}a_{{\bar{q}}_{Fc}^{(\pm)},c,{1\over 2}}
a_{{\bar{q}}_{Fs}^{(\pm)},s,-{1\over 2}} \, ,\hspace{1cm}
\tilde{c}_{\pm k_{F\downarrow },\downarrow } =
a_{{\bar{q}}_{Fc}^{(\pm)},c,{1\over 2}}V^{\pm}_{s}
a^{\dag }_{q_{s}^{(\pm)},s,-{1\over 2}} \, ,
\end{equation}
for the holes.

The expressions for the $(1,1)$ sector are

\begin{equation}
\tilde{c}^{\dag }_{\pm k_{F\uparrow },\uparrow } =
a^{\dag }_{q_{Fc}^{(\pm)},c,{1\over 2}}V^{\mp}_{s}
a^{\dag }_{q_{s}^{(\pm)},s,{1\over 2}} \, ,\hspace{1cm}
\tilde{c}^{\dag }_{\pm k_{F\downarrow },\downarrow } =
V^{\mp}_{c}a^{\dag }_{q_{Fc}^{(\pm)},c,{1\over 2}}
a^{\dag }_{q_{Fs}^{(\pm)},s,{1\over 2}} \, ,
\end{equation}
for the electrons and

\begin{equation}
\tilde{c}_{\pm k_{F\uparrow },\uparrow } =
a_{{\bar{q}}_{Fc}^{(\pm)},c,{1\over 2}}V^{\pm}_{s}
a_{q_{s}^{(\pm)},s,{1\over 2}} \, ,\hspace{1cm}
\tilde{c}_{\pm k_{F\downarrow },\downarrow } =
V^{\pm}_{c}a_{{\bar{q}}_{Fc}^{(\pm)},c,{1\over 2}}
a_{{\bar{q}}_{Fs}^{(\pm)},s,{1\over 2}} \, ,
\end{equation}
for the holes.

According to Eqs. $(41)-(48)$ the $\sigma $ quasiparticles and
quasiholes are many-pseudohole objects which recombine the 
colors $c$ and $s$ (charge and spin in the 
limit $m = n_{\uparrow}-n_{\downarrow} \rightarrow 0$ 
\cite{Carmelo94c,Carmelo95b}) 
giving rise to spin projection $\uparrow $ and $\downarrow $
and have Fermi surfaces at $\pm k_{F\sigma }$. 

Similar expressions can be derived for the sectors of
parameter space where the Hamiltonian $(1)$ has higher
symmetry. We start by considering the sectors of
Hamiltonian symmetry $SU(2)\otimes U(1)$ where

\begin{equation}
q_{Fc}^{(+)}=-q_{Fc}^{(-)}=
q_{c}^{(+)}=-q_{c}^{(-)}=
\pi[1-{1\over {N_a}}] \, .
\end{equation}
For ground states of the $l'=-1$ sector of Hamiltonian 
symmetry $SU(2)\otimes U(1)$ the electrons read

\begin{equation}
\tilde{c}^{\dag }_{\pm k_{F\uparrow },\uparrow } =
V^{\mp}_{c}a^{\dag }_{q_{Fc}^{(\pm)},c,{1\over 2}}
a^{\dag }_{q_{Fs}^{(\pm)},s,-{1\over 2}} \, ,\hspace{1cm}
\tilde{c}^{\dag }_{\pm k_{F\downarrow },\downarrow } =
a^{\dag }_{q_{Fc}^{(\pm)},c,{1\over 2}}V^{\mp}_{s}
a_{q_{s}^{(\pm)},s,-{1\over 2}} \, ,
\end{equation}
and the holes read

\begin{equation}
\tilde{c}_{\pm k_{F\uparrow },\uparrow } =
a^{\dag }_{q_{Fc}^{(\pm)},c,-{1\over 2}}V^{\mp}_{s}
a_{q_{s}^{(\pm)},s,-{1\over 2}} \, ,\hspace{1cm}
\tilde{c}_{\pm k_{F\downarrow },\downarrow } =
V^{\mp}_{c}a^{\dag }_{q_{Fc}^{(\pm)},c,-{1\over 2}}
a^{\dag }_{q_{Fs}^{(\pm)},s,-{1\over 2}} \, .
\end{equation}

For the $l'=1$ sector of Hamiltonian symmetry $SU(2)\otimes U(1)$ 
the electrons read

\begin{equation}
\tilde{c}^{\dag }_{\pm k_{F\uparrow },\uparrow } =
a^{\dag }_{q_{Fc}^{(\pm)},c,{1\over 2}}V^{\mp}_{s}
a^{\dag }_{q_{s}^{(\pm)},s,{1\over 2}} \, ,\hspace{1cm}
\tilde{c}^{\dag }_{\pm k_{F\downarrow },\downarrow } =
V^{\mp}_{c}a^{\dag }_{q_{Fc}^{(\pm)},c,{1\over 2}}
a^{\dag }_{q_{Fs}^{(\pm)},s,{1\over 2}} \, ,
\end{equation}
and the holes read

\begin{equation}
\tilde{c}_{\pm k_{F\uparrow },\uparrow } =
V^{\mp}_{c}a^{\dag }_{q_{Fc}^{(\pm)},c,-{1\over 2}}
a^{\dag }_{q_{Fs}^{(\pm)},s,{1\over 2}} \, ,\hspace{1cm}
\tilde{c}_{\pm k_{F\downarrow },\downarrow } =
V^{\mp}_{s}a^{\dag }_{q_{Fc}^{(\pm)},c,-{1\over 2}}
a_{q_{s}^{(\pm)},s,{1\over 2}} \, .
\end{equation}

In the sectors of Hamiltonian symmetry $U(1)\otimes SU(2)$ we
have that

\begin{equation}
q_{Fs}^{(+)}=-q_{Fs}^{(-)}=
q_{s}^{(+)}=-q_{s}^{(-)}=
\pi [{n\over 2}-{1\over {N_a}}] \, .
\end{equation}
In the case of the $l=-1$ sector of Hamiltonian symmetry 
$U(1)\otimes SU(2)$ the up-spin electron and hole read

\begin{equation}
\tilde{c}^{\dag }_{\pm k_{F\uparrow },\uparrow } =
V^{\pm}_{c}a_{{\bar{q}}_{Fc}^{(\pm)},c,-{1\over 2}}
a^{\dag }_{q_{s}^{(\pm)},s,{-1\over 2}} \, ,\hspace{1cm}
\tilde{c}_{\pm k_{F\uparrow },\uparrow } =
V^{\mp}_{c}a^{\dag }_{q_{Fc}^{(\pm)},c,-{1\over 2}}
a^{\dag }_{q_{s}^{(\pm)},s,{1\over 2}} \, ,
\end{equation}
and the down-spin electron and hole read

\begin{equation}
\tilde{c}^{\dag }_{\pm k_{F\downarrow },\downarrow } =
V^{\pm}_{c}a_{{\bar{q}}_{Fc}^{(\pm)},c,-{1\over 2}}
a^{\dag }_{{\bar{q}}_{s}^{(\pm)},s,{1\over 2}} \, ,\hspace{1cm}
\tilde{c}_{\pm k_{F\downarrow },\downarrow } =
V^{\mp}_{c}a^{\dag }_{q_{Fc}^{(\pm)},c,-{1\over 2}}
a^{\dag }_{q_{s}^{(\pm)},s,-{1\over 2}} \, .
\end{equation}

For the $l=1$ the sector of Hamiltonian symmetry 
$U(1)\otimes SU(2)$ the up-spin electron and hole read

\begin{equation}
\tilde{c}^{\dag }_{\pm k_{F\uparrow },\uparrow } =
V^{\mp}_{c}a^{\dag }_{q_{Fc}^{(\pm)},c,{1\over 2}}
a^{\dag }_{q_{s}^{(\pm)},s,-{1\over 2}} \, , \hspace{1cm}
\tilde{c}_{\pm k_{F\uparrow },\uparrow } =
V^{\pm}_{c}a_{{\bar{q}}_{Fc}^{(\pm)},c,{1\over 2}}
a^{\dag }_{q_{s}^{(\pm)},s,{1\over 2}} \, ,
\end{equation}
and the down-spin electron and hole read

\begin{equation}
\tilde{c}^{\dag }_{\pm k_{F\downarrow },\downarrow } =
V^{\mp}_{c}a^{\dag }_{q_{Fc}^{(\pm)},c,{1\over 2}}
a^{\dag }_{q_{s}^{(\pm)},s,{1\over 2}} \, ,\hspace{1cm}
\tilde{c}_{\pm k_{F\downarrow },\downarrow } =
V^{\pm}_{c}a_{{\bar{q}}_{Fc}^{(\pm)},c,{1\over 2}}
a^{\dag }_{q_{s}^{(\pm)},s,{-1\over 2}} \, .
\end{equation}

Finally, for the $SO(4)$ initial ground state both Eq. $(49)$
and the following equation

\begin{equation}
q_{Fs}^{(+)}=-q_{Fs}^{(-)}=
q_{s}^{(+)}=-q_{s}^{(-)}=
\pi [{1\over 2}-{1\over {N_a}}] \, ,
\end{equation}
hold true and we find for the electrons

\begin{equation}
\tilde{c}^{\dag }_{\pm k_{F\uparrow },\uparrow } =
V^{\mp}_{c}a^{\dag }_{q_{Fc}^{(\pm)},c,{1\over 2}}
a^{\dag }_{q_{Fs}^{(\pm)},s,-{1\over 2}} \, ,\hspace{1cm}
\tilde{c}^{\dag }_{\pm k_{F\downarrow },\downarrow } =
V^{\mp}_{c}a^{\dag }_{q_{Fc}^{(\pm)},c,{1\over 2}}
a^{\dag }_{q_{Fs}^{(\pm)},s,{1\over 2}} \, ,
\end{equation}
and for the holes

\begin{equation}
\tilde{c}_{\pm k_{F\uparrow },\uparrow } =
V^{\mp}_{c}a^{\dag }_{q_{Fc}^{(\pm)},c,-{1\over 2}}
a^{\dag }_{q_{Fs}^{(\pm)},s,{1\over 2}} \, ,\hspace{1cm}
\tilde{c}_{\pm k_{F\downarrow },\downarrow } =
V^{\mp}_{c}a^{\dag }_{q_{Fc}^{(\pm)},c,-{1\over 2}}
a^{\dag }_{q_{Fs}^{(\pm)},s,-{1\over 2}} \, .
\end{equation}

Equations $(60)-(61)$ reveal that removing or adding electrons from
the $SO(4)$ ground state always involves creation
of pseudoholes. Furthermore, while in the
case of the $(l,l')$ sectors the initial and
final ground states belong in general to the same sector,
in the case of the $SO(4)$ ground state each of the four 
possible transitions associated with adding one
up-spin or one down-spin quasiparticle or quasihole leads to
four ground states belonging to a different
$(l,l')$ sector. If the initial ground state belongs
to the $(l')$ $SU(2)\otimes U(1)$ sector [or to the  
$(l)$ $U(1)\otimes SU(2)$ sector] then two
of the final ground states belong to the $(1,l')$
[or to the $(l,1)$] sector and the remaining two
ground states to the $(-1,l')$ [or to the $(l,-1)$] sector.

Equations $(32)$ and $(33)$ tell us that the values of $\eta $ and
$\eta_z$ are fully determined by the number of $c,\beta$ pseudoholes
whereas the number of $s,\beta$ pseudoholes determines the values of
$S$ and $S_z$. In addition, note that the quasiparticle and quasihole 
operators $(41)-(48)$, $(51)-(53)$, $(55)-(58)$, and $(60)-(61)$
involve always a change in the number of $c$ 
pseudoholes of one and a change in the number of $s$ pseudoholes 
also of one. Moreover, when acting on the suitable ground state 
these operators change the values of $\eta $ and $\eta_z$ by 
$\pm 1/2$ and $\pm sgn (\eta_z)1/2$, respectively, and the values 
of $S$ and $S_z$ by $\pm 1/2$ and $\pm sgn (S_z)1/2$, respectively. 
(The corresponding changes in the pseudohole numbers and in the 
values of $\eta $, $\eta_z$, $S$, and $S_z$ are shown in Tables II 
and III.) The analysis of the changes in the pseudohole numbers 
could lead to the conclusion that the $c,\pm {1\over 2}$ pseudoholes have 
quantum numbers $(\eta =1/2;S=0;\eta_z=\pm 1/2;S_z=0)$ and that 
the $s,\mp {1\over 2}$ pseudoholes have quantum numbers 
$(\eta =0;S=1/2;\eta_z =0;S_z=\pm 1/2)$. 
If this was true the $c$ and corresponding $\beta$ pseudohole quantum 
numbers could be identified with $\eta $ and $\eta_z$, respectively, 
and the $s$ and corresponding $\beta$ pseudohole quantum numbers could be 
identified with $S$ and $S_z$, respectively. However, this is not in 
general true. This holds true in the particular case of 
zero-momentum number operators. On the other hand, the above
identities are also true for finite-momentum operators
for $c,\pm {1\over 2}$ in the limit of zero chemical 
potential and for $s,\pm {1\over 2}$ in the limit of zero magnetic 
field, as we find below. 

In order to confirm that the above equivalences are not in 
general true for finite-momentum fluctuations we consider 
the $\alpha $-pseudohole fluctuation operator 

\begin{equation}
\rho_{\alpha }(k) =
- \sum_{q,\beta}\beta a_{q,\alpha ,\beta}^{\dag }a_{q+k,\alpha,\beta} 
\, , 
\end{equation}
and the $\eta_z$- (charge) and $S_z$- (spin) fluctuation
operators 

\begin{equation}
\rho_{\eta_z }(k) =\sum_{k',\sigma}
\left[{1\over 2}\delta_{k,o} - 
c_{k'+k,\sigma}^{\dag }c_{k',\sigma}\right] \, , 
\end{equation}
and

\begin{equation}
\rho_{S_z }(k) =\sum_{k',\sigma}\sigma c_{k'+k,\sigma}^{\dag }
c_{k',\sigma} \, ,  
\end{equation}
respectively. From equations $(32)$ and $(33)$ we find

\begin{equation}
\rho_{c}(0) =\rho_{\eta_z }(0) = N_a - N_{\uparrow }
- N_{\downarrow } \, , 
\end{equation}
and

\begin{equation}
\rho_{s}(0) = \rho_{S_z }(0) = N_{\uparrow }
- N_{\downarrow } \, .  
\end{equation}
We then conclude that at zero momentum the above
equivalences hold true. The electron numbers $N_{\uparrow }$ and 
$N_{\downarrow }$ are good quantum numbers of the 
many-electron system. Since the exact Hamiltonian eigenstates are 
simple Slater determinants of $\alpha ,\beta$-pseudohole levels,
the numbers of $\alpha ,\beta$ pseudoholes are thus required to be 
also good quantum numbers. They are such that Eqs. $(65)$ 
and $(66)$ are obeyed. 

On the other hand, the conservation of electron and
pseudohole numbers does not require the finite-momentum
$c$ and $s$ fluctuations being bare finite-momentum charge and 
spin fluctuations, respectively. By simplicity, we consider 
the smallest momentum values, $k\pm {2\pi\over {N_a}}$.
We emphasize that for $k=\pm {2\pi\over {N_a}}$, acting 
the operator $\rho_{\alpha }(k)$ onto a ground state 
of general form $(17)$ generates a sinlge-pair 
$\alpha $-pseudoparticle-pseudohole excitation where the 
$\alpha ,\beta$ pseudohole at $q={\bar{q}}_{F\alpha }^{(\pm)}$ moves to 
$q=q_{F\alpha }^{(\pm)}$ [see Eq. $(18)$]. If in $c,\beta$
the color $c$ was eta spin and $\beta=\eta_z $ and in $s,\beta$ the
color $s $ was spin and $\beta=S_z $, we should have
that $\rho_{c}(\pm {2\pi\over {N_a}})=\rho_{\eta_z }
(\pm {2\pi\over {N_a}})$ and $\rho_{s}(\pm {2\pi\over 
{N_a}})=\rho_{S_z }(\pm {2\pi\over {N_a}})$, respectively. 
However, the results of Ref.
\cite{Carmelo94c} show that this is not true for the
sectors of Hamiltonian symmetry $U(1)\otimes U(1)$. Altough the
pseudohole summations of Eqs. $(32)$ and $(33)$ give $\eta$, 
$S$, $\eta_z$, and $S_z$ this does not require each $c$
pseudohole having eta spin $1/2$ and spin $0$ and each $s$
pseudohole having eta spin $0$ and spin $1/2$. Also, the fact 
that the quasiparticle or quasihole of Eqs. $(41)-(48)$, $(50)-(53)$, 
$(55)-(58)$, and $(60)-(61)$ has 
$\eta =1/2;S=1/2;\eta_z=sgn (\eta_z)1/2;
S_z=sgn (S_z)1/2$ does not tell how these
values are destributed by the corresponding 
$c$ pseudohole, $s$ pseudohole, and topological momenton.

The studies of Ref. \cite{Carmelo94c} reveal that for finite 
values of the chemical potential and magnetic field there is 
a $c$ and $s$ separation of the low-energy and small-momentum 
excitations but that the orthogonal modes $c$ and $s$ are not 
in general charge and spin, respectively \cite{Carmelo94c}. On 
the other hand, in that reference it was found that in the limit 
of zero chemical potential the finite-momentum $c$ fluctuations 
become real charge excitations and that in the limit of zero 
magnetic field the finite-momentum $s$ fluctuations become real 
spin excitations. In the latter limit the $c$ excitations
are also real charge excitations and the $c$ and $s$
low-energy separation becomes the usual charge and
spin separation \cite{Solyom,Voit,Metzner,Anderson}. In these
limits $c,\beta$ becomes $\eta ,\eta_z$ and $s,\beta$ becomes
$S,-S_z$.

It follows that in the case of the $SO(4)$ canonical ensemble 
the set of pseudoholes involved in the description of the two 
electron and two hole operators $(60)-(61)$,
which are the $c,+{1\over 2}$; $c,-{1\over 2}$; 
$s,+{1\over 2}$; and $s,-{1\over 2}$ pseudoholes 
at the pseudo-Fermi points, transform
in the $\eta =1/2$ and $S=1/2$ representation of the $SO(4)$
group. Moreover, it can be shown from the changes in the BA quantum
numbers and from the study of the pseudohole energies
that the $\eta =1/2$ and $S=1/2$ elementary
excitations studied in Ref. \cite{Essler} are simple combinations 
of one of the ground-state -- ground-state transitions
generated by the operators $(60)-(61)$ with a single
pseudoparticle-pseudohole process relative to the final
ground state. In addition, the usual half-filling holons and 
zero-magnetization spinons can be shown to be limiting cases 
of our pseudohole excitations. For instance, the  
$(\eta =1/2;S=0;\eta_z=1/2;S_z =0)$ anti holon and
$(\eta =1/2;S=0;\eta_z=-1/2;S_z =0)$ holon excitations of Ref. 
\cite{Essler} are at lowest energy generated from the $SO(4)$ ground state
by the operators $V^{-}_{c}
a^{\dag }_{q_{Fc}^{(\pm)},c,{1\over 2}}$ and 
$V^{+}_{c}a^{\dag }_{q_{Fc}^{(\pm)},c,-{1\over 2}}$, 
respectively. Also at lowest energy, the two 
$(\eta =0;S=1/2;\eta_z =0;S_z=1/2)$ and 
$(\eta =0;S=1/2;\eta_z =0;S_z=-1/2)$ spinons
\cite{Essler} are generated from that ground state by the operators 
$a^{\dag }_{q_{Fs}^{(\pm)},s,-{1\over 2}}$ and $a^{\dag 
}_{q_{Fs}^{(\pm)},s,{1\over 2}}$, respectively. The full spectrum
of these excitations is obtained by adding to these generators
a suitable single pseudoparticle-pseudohole-pair operator.
The corresponding energy spectrum involves the pseudohole
bands (A17) and (A18) and by use of the momentum
expressions $(29)-(31)$ and Hamiltonian expression
(A15) recovers in the limit of zero chemical potential
and magnetic field the expressions of Ref. \cite{Essler}.
Therefore, our expressions (B1) and $(60)-(61)$ define the electron
in terms of holons and spinons. Since only Hamiltonian
eigenstates with integer values of $\eta_z +S_z$ are allowed,
the holon -- spinon pairs of Eqs. $(60)-(61)$ cannot be separated.
This also holds true in the general case, the electron
being constituted by one $c$ pseudohole, one $s$ pseudohole, 
and one many-pseudohole topological momenton of large
momentum, as confirmed by Eqs. $(41)-(48)$, $(50)-(53)$,
and $(55)-(58)$. Also in this case the fact that only 
Hamiltonian eigenstates with integer values of 
$\eta_z +S_z$ are allowed prevents these three excitations
of being separated. 

As for the $SO(4)$ ground state, we can relate the symmetry
of the Hamiltonian $(1)$ in a given canonical ensemble by 
looking at the pseudohole contents of the corresponding two 
electrons and two holes of Eqs. $(41)-(48)$, $(50)-(53)$, and
$(55)-(58)$. For instance, Eqs. 
$(41)-(48)$ show that in the $(l,l')$ sectors of Hamiltonian 
symmetry $U(1)\otimes U(1)$ the two electrons and two holes  
involve one pair of the same type of pseudoholes, namely the 
corresponding $c,{l\over 2}$ and $s,{l'\over 2}$ pseudoholes. 
Each of these transforms in the representation of the group 
$U(1)$ and, therefore, the set of two pseudoholes transforms 
in the representation of the group $U(1)\otimes U(1)$.

In the case of the $(l')$ sectors of Hamiltonian symmetry 
$SU(2)\otimes U(1)$ $c$ is eta spin and the corresponding
quantum number $\beta$ is $\eta_z$ and 
Eqs. $(50)-(53)$ confirm that the two electrons and holes involve 
either one $c,{1\over 2}$ pseudohole or one $c,-{1\over 2}$ 
pseudohole combined with one $s,{l'\over 2}$ pseudohole. 
The $c,{1\over 2}$ and $c,-{1\over 2}$ pseudoholes transform in 
the $\eta =1/2$ representation of the eta-spin $SU(2)$ group, whereas 
the $s,{l'\over 2}$ pseudohole transforms in the representation of the 
$U(1)$ group. Therefore, the set of $c,{1\over 2}$; 
$c,-{1\over 2}$; and $s,{l'\over 2}$ pseudoholes transforms 
in the $\eta =1/2$ representation of the $SU(2)\otimes U(1)$ group. 

In the case of the $(l)$ sectors of Hamiltonian symmetry 
$U(1)\otimes SU(2)$ $s$ is spin and the corresponding
quantum number $\beta$ is $\beta=S_z$ and Eqs. $(55)-(58)$ 
show that the two electrons and two holes are constituted by either 
one $s,{1\over 2}$ or one $s,-{1\over 2}$ pseudohole combined 
with one $c,{l\over 2}$ pseudohole. The $s,{1\over 2}$ and 
$s,-{1\over 2}$ pseudoholes transform in the $S =1/2$ 
representation of the spin $SU(2)$ group and the $c,{l\over 2}$ 
pseudohole transforms in the representation of the $U(1)$ group. It
follows that the set of the $c,{l\over 2}$; $s,{1\over 2}$; 
and $s,-{1\over 2}$ pseudoholes transforms in the $S =1/2$ 
representation of the $U(1)\otimes SU(2)$ group.

%%%%%%%%%%%%%%%%%%%%%%%%%%%%%%%%%%%%%%%%%%%%%%%%%%%%%%%%%%%%%%%%%
\section{CONCLUDING REMARKS}

In this paper we have introduced a pseudohole representation
for the states I of the Hubbard chain in a magnetic field
and chemical potential which is valid for all sectors of 
Hamiltonian symmetry. In the pseudohole picture {\it all}
Hamiltonian eigenstates can be generated from a single pseudohole
reference vacuum, the half-filling and zero-magnetic field
ground state. This differs from the pseudoparticle description
of Ref. \cite{Carmelo95} which requires four different
reference vacua.

The introduction of the above pseudohole description
has allowed the study of the interplay between Hamiltonian symmetry 
in each of the nine sectors of parameter space and
the transformation laws of the set of pseudoholes which form 
the electrons and holes of vanishing excitation energy.
This study has required the generalization of the $(-1,-1)$- 
sector results of Ref. \cite{Carmelo95b} to all sectors of 
parameter space. For all the nine sectors we could 
express the Fermi-momentum $\pm k_{F\sigma }$ electrons and holes 
in terms of one pair of pseudoholes and one topological 
momenton. These three quantum objects are confined in
the electron or hole and cannot be separated. 
We have considered the particular
set of the up-spin electron, down-spin electron, 
up-spin hole, and down-spin hole whose individual addition 
to an initial ground state leads to the four final ground states 
differing from it by one electron number. We found that the
two, three, or four different types of $\alpha ,\beta$ pseudoholes 
which are contained (two in each electron or hole) in that set 
always transform in the representation of the symmetry group of 
the Hamiltonian in the sector of parameter space of the 
initial ground state. 

We have also shown that the usual half-filling holons
and zero-magnetization spinons are limiting cases of
our pseudohole and topological momenton excitations. 
Our operator study has allowed the identification of
the holon and spinon generators as well as the exact holon 
and spinon contents of the $SO(4)$ electrons and holes
of vanishing excitation energy.

Finally, we will consider elsewhere an extension of the present 
pseudohole basis which refers to the whole Hilbert space of the
Hamiltonian $(1)$. In addition to the 
$\alpha ,\beta$ pseudoholes, this requires the
introduction of new branches of ``heavy'' 
pseudoparticles \cite{Carmelo95b,Carmelo96}. 
These heavy pseudoparticles are 
absent in the states I, the construction of the states II 
including their creation onto the pseudohole vacuum. 
Thus the universal pseudohole vacuum introduced in this paper
and the associate pseudohole basis provide the
correct and suitable starting point for the extension
of our operator description to the whole Hilbert space.

Although the $\alpha ,\beta$ pseudoholes (or, equivalentely, the 
$\alpha $ pseudoparticles) associated with the states I are the 
transport carriers at low energy \cite{Carmelo92,Carmelo92b} 
and couple to external potentials \cite{Carmelo93}, they 
refer to purely non-dissipative excitations, {\it i.e.} the 
Hamiltonian $(1)$ {\it commutes} with the charge current operator
in the Hilbert subspace spanned by the states I \cite{Carmelo92b}. 
Therefore, the pseudohole currents give 
rise {\it only} to the coherent part of the conductivity
spectrum, i.e. to the Drude peak \cite{Carmelo92b}. 
The finite-frequency part is associated with the above
``heavy'' pseudoparticles. For instance, we will show 
elsewhere \cite{Carmelo95b,Carmelo96} that both the 
$c,\beta $ pseudoholes and some of the heavy pseudoparticles
couple to external vector potentials in such a way that 
in the full Hilbert space the Hamiltonian {\it does not} 
commute with the charge current operator.

%%%%%%%%%%%%%%%%%%%%%%%%%%%%%%%%%%%%%%%%%%%%%%%%%%%%%%%%%%%%%%%%%%%
\nonum
\section{ACKNOWLEDGMENTS}

This work was supported in part by the Institute for Scientific 
Interchange Foundation (Torino). We thank F. Essler, G. Kotliar, 
and A. H. Castro Neto for stimulating discussions.

%%%%%%%%%%%%%%%%%%%%%%%%%%%%%%%%%%%%%%%%%%%%%%%%%%%%%%%%%%%%%%%%%%%
\vfill
\eject
\appendix{THE PSEUDOHOLE BASIS AND THE BA SOLUTION}

In this Appendix we discuss the pseudohole and pseudoparticle
descriptions and relate the pseudohole number operator 
$\hat{N}^h_{\alpha ,\beta}(q)=a^{\dag }_{q,\alpha ,\beta}
a_{q,\alpha ,\beta}$ to the BA equations. We also present the 
Hamiltonian in the pseudohole basis and the associate disperson 
relations. 

In order to relate the present pseudoholes to the pseudoparticle
description of Ref. \cite{Carmelo95}, we emphasize that the 
$\alpha (l,l')$ pseudoholes associated with the pseudoparticles 
introduced in that reference are such that $c(l,1)=c(l,-1)$ and 
$s(1,l')=s(-1,l')$. This is related to the fact that 
for states I with the same values 
of $\eta $ and $S$ the $\alpha (l,l')$ pseudomomentum-orbital 
numbers $N_{\alpha (l,l')}^*$ [and $\alpha (l,l')$ pseudoparticle 
numbers $N_{\alpha (l,l')}$] shown in Table I of Ref. \cite{Carmelo95} 
are for different $(l,l')$ numbers equal. This refers to pairs of 
states I where one is a LWS and the other is the corresponding 
HWS of the same family of multiplets of eta-spin or spin algebras. 
In this way these numbers are $l$- and $l'$-independent [see Eq. 
$(9)$], ie do not depend on the signs of $\eta_z$ and $S_z$ but 
only on the corresponding values of $\eta$ and $S$. 
Therefore, we can refer them simply by $N_{\alpha }^*$ and 
$N_{\alpha }$. Their general $\eta$- and $S$-dependent 
expressions are given in Eqs. $(10)$ and $(11)$. This does 
not affect the conclusions and results of Ref. 
\cite{Carmelo95} which remain fully
correct. The only consequence is the simplifying reduction of 
the problem to four pseudohole branches which we denote in 
general by $\alpha ,\beta$ pseudoholes. The colors $c$ 
and $s$ and quantum numbers $\beta=\pm {1\over 2}$ which label the
four pseudohole branches also label the Hamiltonian eigenstates
I, as discussed in Sec. II. Following that section, the description 
of the non-LWS's and (or) non-HWS's multiplets generated from 
the states I \cite{Carmelo95c} reveals that the 
pseudoparticles associated with the $\alpha ,\beta$ pseudoholes 
should simply be denoted as $\alpha $ pseudoparticles {\it and 
not} as $\alpha ,\beta$ pseudoparticles. Therefore, the $\alpha $ 
pseudoparticle operators obey the following 
anticommuting algebra which does not include the pseudohole 
quantum number $\beta$:

\begin{equation}
\{b^{\dag }_{q,\alpha },b_{q' ,\alpha'}\} 
= \delta_{q,q'}\delta_{\alpha ,\alpha'}\, , 
\end{equation}
and

\begin{equation}
\{b^{\dag }_{q,\alpha},b^{\dag }_{q',\alpha'}\} 
= \{b_{q,\alpha},b_{q',\alpha'}\} 
= 0 \, . 
\end{equation}

Moreover, note that in the present case of the states I and $(l,l')$ 
sectors of parameter space 
the following selection rule is valid: out of the four $\alpha 
,\beta$ pseudohole branches only the two branches $c,{l\over 2}$ 
and $s,{l'\over 2}$ contribute to the generators of Eq. $(17)$. 
(The states I are constructed by acting these generators onto 
the pseudohole vacuum.) Since in this case there is in the
$\alpha $ orbital single occupancy by one of the two $\beta$ 
pseudohole branches only, one could denote the associate $\alpha $ 
pseudoparticles by $\alpha, \beta$ pseudoparticles, with the 
fixed $\beta$ value ${l\over 2}$ or ${l'\over 2}$ for
$c$ or $s$, respectively. 
(These $\alpha, \beta$ pseudoparticles are the {\it holes}
of the corresponding $\alpha, \beta$ pseudoholes -- however,
when there is in the same $\alpha $ orbital occupation of both
$\alpha, {1\over 2}$ and $\alpha, -{1\over 2}$ pseudoholes
that notation is not allowed and their holes are to be denoted by 
$\alpha $ pseudoparticles.) This together with the fact that 
our pseudoholes are related to the $\alpha (l,l')$ pseudoholes 
of Ref. \cite{Carmelo95} as $c,\beta=c(2\beta,1)=c(2\beta,-1)$ 
and $s,\beta =s(1,2\beta)=s(-1,2\beta)$ justifies the 
pseudoparticle notation of Ref. \cite{Carmelo95} which refers 
to states I only. 

Let us denote by $\cal {H}_I$ the Hilbert subspace spanned by 
the Hamiltonian eigenstates I. The above selection rule for 
$\cal {H}_I$ and $(l,l')$ sectors simplifies the $\beta $ summations 
which have only contributions from the ${l\over 2}$ (for $\alpha =c$) 
and ${l'\over 2}$ (for $\alpha =s$) $\beta $ values. For instance, 
the Hamiltonian expression involves in $\cal {H}_I$ the 
pseudomomentum distribution operator 

\begin{equation}
\hat{N}^h_{\alpha}(q)=\sum_{\beta}\hat{N}^h_{\alpha ,\beta}(q)
=\sum_{\beta} a^{\dag }_{q,\alpha ,\beta}a_{q,\alpha ,\beta} \, .
\end{equation}
It has the same information as the corresponding pseudoparticle operator

\begin{equation}
\hat{N}_{\alpha}(q) = 1 - \hat{N}^h_{\alpha}(q) =
b^{\dag }_{q,\alpha }b_{q,\alpha } \, .
\end{equation}
The $\cal {H}_I$ and $(l,l')$-sector selection rule allows 
Eq. (A3) to be simplified to

\begin{equation}
\hat{N}^h_{c}(q) = \hat{N}^h_{c,{l\over 2}}(q) =
a^{\dag }_{q, c,{l\over 2}}a_{q,c,{l\over 2}} \, ;
\hspace{1cm}
\hat{N}^h_{s}(q) = \hat{N}^h_{s,{l'\over 2}}(q) = 
a^{\dag }_{q, s,{l'\over 2}}a_{q,s,{l'\over 2}} \, .
\end{equation}
 
Equations $(10)$, $(11)$, and $(16)$ can be shown to
refer to a larger Hilbert space than $\cal {H}_I$, which
is spanned both by the states I and all their associate
non-LWS's and non-HWS's. In the present case
of $\cal {H}_I$ these equations can be replaced by  

\begin{equation}
N_c^* = N_a \, , \hspace{1cm} N_s^* = 
{1\over 2}\left[N_a - 2(|\eta_z| -|S_z|)\right] \, ,
\end{equation}

\begin{equation}
N_c = N_a - 2|\eta_z| \, , \hspace{1cm} N_s = 
{1\over 2}\left[N_a - 2(|\eta_z| +|S_z|)\right] \, ,
\end{equation}
and

\begin{equation}
N_c^h = 2|\eta_z| \, , \hspace{1cm} N_s^h = 2|S_z| \, ,
\end{equation}
respectively. The operator $\hat{N}^h_{\alpha}$ can be written
in terms of the pseudohole number operator $\hat{N}^h_{\alpha}(q)$
as follows

\begin{equation}
\hat{N}^h_{\alpha} = \hat{N}^*_{\alpha} 
- \hat{N}_{\alpha} = \sum_{q}\hat{N}^h_{\alpha}(q) \, .
\end{equation}

The pseudohole basis of the $(l,l')$ sectors is 
constructed from the BA solution presisely as in Ref.
\cite{Carmelo94} for the particular case of the 
$(-1,-1)$ sector. Two differences are that (i) the
$(-1,-1)$ numbers $N_{\alpha}^*$ and $N_{\alpha}$ of 
Eqs. $(10)$ of Ref. \cite{Carmelo94} are here to be replaced
by the general expressions (A6) and (A7); and (ii) we 
use here pseudoholes instead of pseudoparticles.

The operators (A3) commute with each other, i.e.
$[\hat{N}^h_{\alpha}(q),\hat{N}^h_{\alpha'}(q')]=0$. 
As in the $(-1,-1)$ case \cite{Carmelo94}, in the pseudohole 
basis the $(l,l')$-sector Hamiltonian expression involves the 
operator (A3) [or (A4)]. Furthermore, the Hamiltonian 
{\it commutes} in $\cal {H}_I$ with that 
operator. This plays a central role in this Hilbert subspace 
because all the Hamiltonian eigenstates which are LWS's I or 
HWS's I are also eigenstates of $\hat{N}^h_{\alpha}(q)$. 
Lets us denote such states I by $|\eta_z,S_z\rangle$, where 
$\eta_z$ and $S_z$ are the eigenvalues which 
characterize the canonical ensemble. As for the LWS's of
the $(-1,-1)$ sector, these LWS's I or (and) HWS's I obey 
eigenvalue equations of the form

\begin{equation}
\hat{N}^h_{\alpha}(q)|\eta_z,S_z\rangle =
N^h_{\alpha}(q)|\eta_z,S_z\rangle \, , 
\end{equation}
where $N^h_{\alpha}(q)$ represents the eigenvalue of the 
operator (A3), which is given by $1$ and $0$ for pseudohole
occupied and empty values of $q$, respectively. 
The Hamiltonian reads

\begin{equation}   
\hat{H} = - \sum_{q}[1 - \hat{N}^h_c(q)] 2t\cos [\hat{K}(q)]
 + [2\mu -U]\hat{\eta }_z + 2\mu_0 H\hat{S}_z \, ,
\end{equation}
where the expressions of the diagonal generators are given in 
Eq. $(33)$. This is the exact expression of the Hamiltonian $(1)$ 
in $\cal {H}_I$. At energy scales smaller than the gaps for the
non-LWS's and non-HWS's multiplets, LWS's II and HWS's II, 
Eq. (A11) gives the exact expression of that Hamiltonian in the full 
Hilbert space. 

Despite its simple appearance, the Hamiltonian (A11)
describes a many-pseudohole problem. The reason is that the 
expression of the rapidity operator $\hat{K}(q)$ in terms of 
the operator $\hat{N}^h_{\alpha}(q)$ contains many-pseudohole
interacting terms. As for the $(-1,-1)$ sector, in all 
the $(l,l')$ sectors the operator $\hat{K}(q)$ and associate 
rapidity operator $\hat{S}(q)$ obey the following two
equations which are valid for any Hamiltonian eigenstate I

\begin{equation}
[\hat{K}(q) - {2\over N_a}\sum_{q'}[1 - \hat{N}^h_s(q')]
\tan^{-1}\Bigl(\hat{S}(q') - (4t/U)\sin 
[\hat{K}(q)]\Bigr)]|\eta_z,S_z\rangle = q|\eta_z,S_z\rangle
\end{equation}
and

\begin{eqnarray}
&  & {2\over N_a}[\sum_{q'}[1 - \hat{N}^h_c(q')]
\tan^{-1}\Bigl(\hat{S}(q) - (4t/U)
\sin [\hat{K}(q')]\Bigr)
\nonumber \\ 
& - & \sum_{q'}[1 - \hat{N}^h_s(q')]\tan^{-1}\Bigl({1\over 2}
\left(\hat{S}(q) - \hat{S}(q')\right)\Bigr)]|\eta_z,S_z\rangle = 
q|\eta_z,S_z\rangle \, .
\end{eqnarray}
These equations fully define the rapidity operators in terms 
of the pseudomomentum distribution operator (A3). We
note that the limits of the pseudo-Brillouin zones of
Eqs. (A12) and (A13) pseudomomentum summations are given in
Eqs. $(22)-(24)$. These limits involve the numbers $N^*_{\alpha}$
whose general expressions for the four $(l,l')$ sectors
are given by Eqs. $(10)$ and (A6). Otherwise 
the general Eqs. (A12) and (A13) have the same form as 
Eqs. $(31)$ and $(32)$ of Ref. \cite{Carmelo94}
for the $(-1,-1)$ sector. 

The normal-ordered Hamiltonian relatively to the suitable
ground state of form $(17)$ reads 

\begin{equation}
:\hat{H}: = \sum_{i=1}^{\infty}\hat{H}^{(i)} \, ,
\end{equation} 
where to second pseudohole scattering order

\begin{eqnarray}
\hat{H}^{(1)} & = & -\sum_{q,\alpha} 
\epsilon^0_{\alpha}(q):\hat{N}^h_{\alpha}(q): 
+ [2\mu -U]\sum_{q,\beta}\beta:\hat{N}^h_{c,\beta}(q): 
+ 2\mu_0 H\sum_{\beta}\beta:\hat{N}^h_{s,\beta}(q): \, ;\nonumber\\
\hat{H}^{(2)} & = & {1\over {N_a}}\sum_{q,\alpha} \sum_{q',\alpha'} 
{1\over 2}f_{\alpha\alpha'}(q,q') 
:\hat{N}^h_{\alpha}(q)::\hat{N}^h_{\alpha'}(q'): \, .
\end{eqnarray}
Note that $:\hat{N}^h_{\alpha}(q): = \sum_{\beta} 
:\hat{N}^h_{\alpha,\beta}(q):$. In the present case of 
states I and $(l,l')$ sectors of parameter space
the $\beta $ selection rule has allowed us to write Eq. (A5)
and also

\begin{equation}
\sum_{\beta}\beta :\hat{N}^h_{c,\beta}(q): =
{l\over 2} :\hat{N}^h_{c,{l\over 2}}: \, ; \hspace{1cm}
\sum_{\beta}\beta :\hat{N}^h_{s,\beta}(q): =
{l'\over 2} :\hat{N}^h_{s,{l'\over 2}}: \, .
\end{equation}

Equation (A15) includes the Hamiltonian 
terms which are {\it relevant} at low energy 
\cite{Carmelo94c,Carmelo93b,Carmelo94,Carmelo94b}. 
Furthermore, it is shown in Ref. \cite{Carmelo94b} [for
the $(-1,-1)$ sector] that at low energy and small momentum the 
only relevant term is the non-interacting term $\hat{H}^{(1)}$. 
This property justifies to the Landau-liquid character of 
the Hamiltonian $(1)$ and plays a key role in the symmetries 
of the critical point.

The expressions of the bands are

\begin{equation}
\epsilon_{c}^0(q) = -2t\cos K^{(0)}(q) +
2t\int_{Q^{(-)}}^{Q^{(+)}}
dk\widetilde{\Phi }_{cc}
\left(k,K^{(0)}(q)\right)\sin k \, ,
\end{equation}
and

\begin{equation}
\epsilon_{s}^0(q) = 2t\int_{Q^{(-)}}^{Q^{(+)}}dk
\widetilde{\Phi }_{cs}
\left(k,S^{(0)}(q)\right) \sin k \, ,  
\end{equation}
respectively. Here 

\begin{equation}
Q^{(\pm)} = K^{(0)}(q_{Fc }^{(\pm)}) \, ,
\end{equation}
$K^{(0)}(q)$ and $S^{(0)}(q)$ are the solutions of Eqs. (A12)
and (A13) for the particular case of the suitable ground state 
$(17)$, and the phase shifts $\tilde{\Phi}_{\alpha\alpha '}$ are 
given by

\begin{equation}
\tilde{\Phi}_{cc}(k,k') = 
\bar{\Phi }_{cc}\left({\sin k\over u},
{\sin k'\over u}\right) \, ; \hspace{1cm}
\tilde{\Phi}_{cs}(k,v') = 
\bar{\Phi }_{cs}\left({\sin k\over u},
v'\right) \, ,
\end{equation}

\begin{equation}
\tilde{\Phi}_{sc}(v,k') = 
\bar{\Phi }_{sc}\left(v,
{\sin k'\over u}\right) \, ; \hspace{1cm}
\tilde{\Phi}_{ss}(v,v') = 
\bar{\Phi }_{ss}\left(v,v'\right) \, ,
\end{equation}
where the phase shifts $\bar{\Phi }_{\alpha\alpha '}$ are
defined by the following integral equations

\begin{equation}
\bar{\Phi }_{cc}\left(x,x'\right) = 
{1\over{\pi}}\int_{{B^{(-)}\over u}}^{{B^{(+)}\over u}}
dy''{\bar{\Phi }_{sc}\left(y'',x'\right) 
\over {1+(x-y'')^2}} \, ,
\end{equation}

\begin{equation}
\bar{\Phi }_{cs}\left(x,y'\right) = 
-{1\over{\pi}}\tan^{-1}(x-y') + {1\over{\pi}}
\int_{{B^{(-)}\over u}}^{{B^{(+)}\over u}}
dy''{\bar{\Phi }_{ss}\left(y'',y'\right) 
\over {1+(x-y'')^2}} \, ,
\end{equation}

\begin{equation}
\bar{\Phi }_{sc}\left(y,x'\right) = 
-{1\over{\pi}}\tan^{-1}(y-x') + \int_{{B^{(-)}\over u}}^{{B^{(+)}\over u}}
dy''G(y,y''){\bar{\Phi }}_{sc}\left(y'',x'\right) \, ,
\end{equation}

\begin{eqnarray}
\bar{\Phi }_{ss}\left(y,y'\right) & =
& {1\over{\pi}}\tan^{-1}({y-y'\over{2}}) -
{1\over{\pi^2}}
\int_{{\sin Q^{(-)}\over u}}^{{\sin Q^{(+)}\over u}}dx''{\tan^{-1} 
(x''-y')\over{1+(y-x'')^2}}
\nonumber \\
& + & \int_{{B^{(-)}\over u}}^{{B^{(+)}\over u}}
dy''G(y,y''){\bar{\Phi }}_{ss}\left(y'',y'\right) \, .
\end{eqnarray}
Here

\begin{equation}
B^{(\pm)}/u = S^{(0)}(q_{Fs}^{(\pm)}) \, , 
\end{equation}
and the kernel $G(y,y')$ reads \cite{Carmelo92b}

\begin{equation}
G(y,y') = - {1\over{2\pi}}\left[{1\over{1+((y-y')/2)^2}}\right]
\left[1 - {1\over 2}
\left(t(y)+t(y')+{{l(y)-l(y')}\over{y-y'}}\right)\right] \, ,
\end{equation}
where

\begin{equation}
t(y) = {1\over{\pi}}\left[\tan^{-1}(y + {\sin Q^{(+)}\over u}) 
- \tan^{-1}(y + {\sin Q^{(-)}\over u})\right]\, ,
\end{equation}
and

\begin{equation}
l(y) = {1\over{\pi}}\left[
\ln (1+(y + {\sin Q^{(+)}\over u})^2) -  
\ln (1+(y + {\sin Q^{(-)}\over u})^2)\right] \, .
\end{equation}

The ``Landau'' $f$ function, $f_{\alpha\alpha'}(q,q')$, has 
universal form in terms of the two-pseudohole phase shifts 
$\Phi_{\alpha\alpha'}(q,q')$ defined below and reads

\begin{eqnarray}
f_{\alpha\alpha'}(q,q') & = & 2\pi v_{\alpha}(q) 
\Phi_{\alpha\alpha'}(q,q')  
+ 2\pi v_{\alpha'}(q') \Phi_{\alpha'\alpha}(q',q) \nonumber \\
& + & \sum_{j=\pm 1} \sum_{\alpha'' =c,s}
2\pi v_{\alpha''} \Phi_{\alpha''\alpha}(jq_{F\alpha''},q)
\Phi_{\alpha''\alpha'}(jq_{F\alpha''},q') \, .
\end{eqnarray}
The two-pseudohole phase shifts can be defined in terms of
the phase shifts $\bar{\Phi }_{\alpha\alpha'}$ as 
follows

\begin{equation}
\Phi_{cc}(q,q') = 
\bar{\Phi }_{cc}\left({\sin K^{(0)}(q)\over u},
{\sin K^{(0)}(q')\over u}\right) \, ,
\end{equation}

\begin{equation}
\Phi_{cs}(q,q') = 
\bar{\Phi }_{cs}\left({\sin K^{(0)}(q)\over u},
S^{(0)}(q')\right) \, ,
\end{equation}

\begin{equation}
\Phi_{sc}(q,q') = 
\bar{\Phi }_{sc}\left(S^{(0)}(q),
{\sin K^{(0)}(q')\over u}\right) \, ,
\end{equation}

\begin{equation}
\Phi_{ss}(q,q') = 
\bar{\Phi }_{ss}\left(S^{(0)}(q),
S^{(0)}(q')\right) \, .
\end{equation}
Finally, the pseudohole group velocity appearing in
the $f$ function expression (A30) is given by 

\begin{equation}
v_{\alpha}(q) = {d\epsilon^0_{\alpha}(q) \over {dq}} \, . 
\end{equation}
In particular, the velocity

\begin{equation}
v_{\alpha}\equiv v_{\alpha}(q_{F\alpha}) \, , 
\end{equation}
plays a determining role at the critical point, representing 
the ``light'' velocities which appear in the conformal-invariant 
expressions \cite{Carmelo94c,Frahm}. 

We emphasize that the phase shifts expressions are the same as
for the $(-1,-1)$ sector (see Ref. \cite{Carmelo92b}), except
that the present limits of pseudo-Brillouin zones and 
pseudo-Fermi points involve in the present general case 
the numbers $(10)$-(A6) and $(11)$-(A7), respectively [see Eqs. 
$(18)-(24)$].

Note that although the expressions for the bands $(34)$ and
$(35)$, rapidity $(36)$, $f$ function $(38)$, velocity $(39)$, phase 
shifts (A18)-(A27) and (A39)-(A42) of Ref. \cite{Carmelo95} are 
absolutly correct they are $l$- and $l'$-independent: these 
expressions can be shown to depend only on $\eta $ and $S$ and 
not on the signs of $\eta_z$ and $S_z$ which for the states I 
determine the values of the numbers $l$ and $l'$ [see Eq. $(9)$]. 
Therefore, the corresponding pseudohole quantities presented
in this Appendix have a simpler form than those of Ref.
\cite{Carmelo95}.

%%%%%%%%%%%%%%%%%%%%%%%%%%%%%%%%%%%%%%%%%%%%%%%%%%%%%%%%%%%%%%%%%%%
\vfill
\eject
\appendix{ELECTRON -- QUASIPARTICLE TRANSFORMATION}

In this Appendix we follow the $(-1,-1)$- sector study of Ref. 
\cite{Carmelo95b} and present a short discussion of
the electron - quasiparticle transformation which
relates the electron operator $c^{\dag }_{k_{F\sigma},\sigma}$
to the quasiparticle operator ${\tilde{c}}^{\dag 
}_{k_{F\sigma},\sigma}$ in the limit of vanishing excitation 
energy. The latter operator is defined by Eq. $(34)$.
To leading order in that energy there is a singular
transformation between these two operators which 
as for the $(-1,-1)$ sector \cite{Carmelo95b} reads

\begin{equation}
\tilde{c}^{\dag }_{\pm k_{F\sigma },\sigma } =
{1\over {\sqrt{Z_{\sigma }}}}
c^{\dag }_{\pm k_{F\sigma},\sigma } \, ,
\end{equation}
where the one-electron renormalization factor is
given by $Z_{\sigma}=\lim_{\omega\to 0}Z_{\sigma}(\omega)$
and $Z_{\sigma}(\omega)$ is the small-$\omega $ 
leading-order term of $|\varsigma_{\sigma}||1-{\partial 
\hbox{Re}\Sigma_{\sigma} 
(\pm k_{F\sigma},\omega)\over {\partial\omega}}|^{-1}$. Here
$\Sigma_{\sigma} (k,\omega)$ is the $\sigma $ self 
energy. We emphasize that Eq. (B1) does not apply to the case
when the starting state is the $SO(4)$ or a
$SU(2)\otimes U(1)$ half-filling ground state.
In this case a similar expression holds true
where $\omega $ is replaced by $\omega -\Delta_c$.
Here $\Delta_c$ is the half-filling Mott-Hubbard gap
\cite{Lieb,Carmelo91}. In Ref. \cite{Carmelo95b} it was 
found that in the Hilbert subspace spanned by the states I 
that self energy is given by

\begin{equation}
\hbox{Re}\Sigma_{\sigma} ( k_{F\sigma},\omega ) =
\omega [1-{\omega^{-1-\varsigma_{\sigma}}\over
{c^{\sigma }_0+\sum_{j=1,2.3,...}c^{\sigma }_j\omega^{4j}}}] 
\, ,
\end{equation}
where $c^{\sigma }_j$ with $j=0,1,2,...$ are constants
and $\varsigma_{\sigma}$ is a non-classical interaction
dependent exponent such that $-1<\varsigma_{\sigma}<-1/2$.
It follows that the function $Z_{\sigma}(\omega)$
is of the form

\begin{equation}
Z_{\sigma}(\omega)=
c^{\sigma }_0 \omega^{1+\varsigma_{\sigma}} \, ,
\end{equation}
and vanishes in the limit of $\omega\rightarrow 0 $. 
Thus, $Z_{\sigma}=0$. Although
expression (B1) is very similar to the corresponding expression
for a Fermi liquid, in the present one-dimensional many-electron
problem there is no overlap between the quasiparticle and the electron, 
in contrast to a Fermi liquid.

The singular electron -- quasiparticle transformation
(B1) maps a non-perturbative electronic quantum problem in
a perturbative quasiparticle problem, the factor 
${1\over {\sqrt{Z_{\sigma }}}}$
being absorbed by that transformation. It maps a 
vanishing-spectral-weight electronic problem onto a 
finite-spectral-weight quasiparticle problem. Following Eq. 
$(34)$, the quasiparticle operator $\tilde{c}^{\dag }_{\pm 
k_{F\sigma },\sigma }$ is the generator which transforms 
the ground state $|0; N_{\sigma}, N_{-\sigma}\rangle$ onto 
$|0; N_{\sigma}+1, N_{-\sigma}\rangle$. Apart from the
factor ${1\over {\sqrt{Z_{\sigma }}}}$
absorbed by the transformation it is a $\sigma $ electron 
of momentum $\pm k_{F\sigma }$. Therefore, in Sec. III
we refer often the quasiparticle by electron.

Similar results hold for the hole and corresponding
quasihole which are related by the singular transformation

\begin{equation}
\tilde{c}_{\pm k_{F\sigma },\sigma } =
{1\over {\sqrt{Z_{\sigma }}}}
c_{\pm k_{F\sigma},\sigma } \, ,
\end{equation}
where $Z_{\sigma }$ is the same as in Eq. (B1). As for
the quasiparticle and the electron, in Sec. III we refer
often the quasihole by hole. In that section we evaluate
the quasiparticle and quasihole expressions in terms of
pseudoholes and topological momentons for all nine sectors
of parameter space.

%****************************************************************
%********************* R E F E R E N C E S **********************
%****************************************************************

\newpage

\centerline{TABLES}
\begin{tabbing}
\sl \hspace{2cm} \= \sl $(-1,-1)$ \hspace{0.3cm} \= 
\sl $(-1,1)$ \hspace{0.3cm} \= \sl $(1,-1)$ 
\hspace{1.5cm} \= \sl $(1,1)$\\ 
(A)\,$P$ \> $\pm 2k_F$ \> $\pm 2k_F$ \> $\pm[2\pi-2k_F]$ 
\> $\pm[2\pi-2k_F]$\\ 
(B)\,$P$ \> $\pm k_{F\uparrow}$ \> $\pm k_{F\uparrow}$ 
\> $\pm k_{F\uparrow}$ \> $\pm k_{F\uparrow}$\\ 
(C)\,$P$ \> $\pm k_{F\downarrow}$ \> $\pm k_{F\downarrow}$ 
\> $\pm k_{F\downarrow}$ \> $\pm k_{F\downarrow}$\\ 
(D)\,$P$ \> $0$ \> $0$ \> $0$ \> $0$ 
\label{tableII}
\end{tabbing}

TABLE I -- Values of the ground-state momentum in the four $(l,l')$ 
sectors of Hamiltonian symmetry $U(1)\otimes U(1)$. The different 
momentum values correspond to the following parities of the numbers 
$N^*_s$ and $N_s$ of Eqs. $(10)$ and $(11)$, respectively: 
(A) even, even; (B) even, odd; (C) odd, even; and (D) 
odd, odd. In the case of the $(1,\pm 1)$ sectors, if 
$k_{F\sigma}>\pi$ then $\pm k_{F\sigma}$ should be replaced
by the first-Brillouin-zone momenta $\pm [2\pi -k_{F\sigma}]$.

\vspace{0.25cm}

\begin{tabbing}
\sl \hspace{3cm} \= \sl $(-1,-1)$ \hspace{0.3cm} \= 
\sl $(-1,1)$ \hspace{0.3cm} \= \sl $(1,-1)$ 
\hspace{0.3cm} \= \sl $(1,1)$\\ 
$\Delta N_{c}^h$ \> $\mp 1$ \> $\mp 1$ \> $\pm 1$ \> $\pm 1$\\ 
$\Delta N_{c}$ \> $\pm 1$ \> $\pm 1$ \> $\mp 1$ \> $\mp 1$\\ 
$\Delta N_{c}^*$ \> $0$ \> $0$ \> $0$ \> $0$\\ 
$\Delta N_{s}^h$ \> $\pm 1$ \> $\mp 1$ \> $\pm 1$ \> $\mp 1$\\ 
$\Delta N_{s}$ \> $0$ \> $\pm 1$ \> $\mp 1$ \> $0$\\ 
$\Delta N_{s}^*$ \> $\pm 1$ \> $0$ \> $0$ \> $\mp 1$\\ 
$\Delta\eta$ \> $\mp 1/2$ \> $\mp 1/2$ \> $\pm 1/2$ \> $\pm 1/2$\\ 
$\Delta\eta_z$ \> $\pm 1/2$ \> $\pm 1/2$ \> $\pm 1/2$ \> $\pm 1/2$\\ 
$\Delta S$ \> $\pm 1/2$ \> $\mp 1/2$ \> $\pm 1/2$ \> $\mp 1/2$\\ 
$\Delta S_z$ \> $\mp 1/2$ \> $\mp 1/2$ \> $\mp 1/2$ \> $\mp 1/2$ 
\label{tableIII}
\end{tabbing}

TABLE II -- Changes in the numbers of pseudoholes, pseudoparticles,
pseudoparticle orbitals and of the values of $\eta$, $\eta_z$, $S$, 
and $S_z$ in the ground-state -- ground-state transition 
$(N_{\uparrow},N_{\downarrow})\rightarrow
(N_{\uparrow}\pm 1,N_{\downarrow})$.

\vspace{0.25cm}

\begin{tabbing}
\sl \hspace{3cm} \= \sl $(-1,-1)$ \hspace{0.3cm} \= 
\sl $(-1,1)$ \hspace{0.3cm} \= \sl $(1,-1)$ 
\hspace{0.3cm} \= \sl $(1,1)$\\ 
$\Delta N_{c}^h$ \> $\mp 1$ \> $\mp 1$ \> $\pm 1$ \> $\pm 1$\\ 
$\Delta N_{c}$ \> $\pm 1$ \> $\pm 1$ \> $\mp 1$ \> $\mp 1$\\ 
$\Delta N_{c}^*$ \> $0$ \> $0$ \> $0$ \> $0$\\ 
$\Delta N_{s}^h$ \> $\mp 1$ \> $\pm 1$ \> $\mp 1$ \> $\pm 1$\\ 
$\Delta N_{s}$ \> $\pm 1$ \> $0$ \> $0$ \> $\mp 1$\\ 
$\Delta N_{s}^*$ \> $0$ \> $\pm 1$ \> $\mp 1$ \> $0$\\ 
$\Delta\eta$ \> $\mp 1/2$ \> $\mp 1/2$ \> $\pm 1/2$ \> $\pm 1/2$\\ 
$\Delta\eta_z$ \> $\pm 1/2$ \> $\pm 1/2$ \> $\pm 1/2$ \> $\pm 1/2$\\ 
$\Delta S$ \> $\mp 1/2$ \> $\pm 1/2$ \> $\mp 1/2$ \> $\pm 1/2$\\ 
$\Delta S_z$ \> $\pm 1/2$ \> $\pm 1/2$ \> $\pm 1/2$ \> $\pm 1/2$ 
\label{tableIV}
\end{tabbing}

TABLE III -- Changes in the numbers of pseudoholes, pseudoparticles,
pseudoparticle orbitals and in the values of $\eta$, $\eta_z$, $S$, 
and $S_z$ in the ground-state -- ground-state transition 
$(N_{\uparrow},N_{\downarrow})\rightarrow
(N_{\uparrow},N_{\downarrow}\pm 1)$.

\begin{references}
\bibitem[1]{Pines}
        D. Pines and P. Nozi\`eres, in {\em The Theory of 
        Quantum Liquids},
        (Addison-Wesley, Redwood City, 1966 and 1989), Vol. I.
\bibitem[2]{Baym}
        Gordon Baym and Christopher J. Pethick, in
        {\em Landau Fermi-Liquid Theory Concepts and Applications},
        (John Wiley \& Sons, New York, 1991).
\bibitem[3]{Solyom}
        J. S\'olyom, Adv. Phys. {\bf 28}, 201 (1979).
\bibitem[4]{Voit}
        V. Meden and K. Sch\"onhammer, Phys. Rev. B
        {\bf 46}, 15 753 (1992);
        J. Voit, Phys. Rev. B {\bf 47}, 6740 (1993).
\bibitem[5]{Metzner}
        Walter Metzner and Carlo Di Castro, Phys. Rev. B 
        {\bf 47}, 16 107 (1993).
\bibitem[6]{Anderson}      
        P. W. Anderson and Y. Ren, in {\it High Temperature
        Superconductivity}, edited by K. S. Bedell,
        D. E. Meltzer, D. Pines, and J. R. Schrieffer
        (Addison-Wesley, Reading, MA, 1990).
\bibitem[7]{Carmelo94c}
        J. M. P. Carmelo, A. H. Castro Neto, and 
        D. K. Campbell, Phys. Rev. Lett. {\bf 73}, 926
        (1994).
\bibitem[8]{Bethe}
        H. A. Bethe, Z. Phys. {\bf 71}, 205 (1931).
\bibitem[9]{Yang}
        For one of the first generalizations of the Bethe
        ansatz to multicomponent systems see
        C. N. Yang, Phys. Rev. Lett. {\bf 19}, 1312
        (1967).
\bibitem[10]{Korepinrev}
        V. E. Korepin, N. M. Bogoliubov, and A. G. Izergin,
        {\it Quantum Inverse Scattering Method and Correlation Functions}
        (Cambridge University Press, 1993).
\bibitem[11]{Lieb}
        Elliott H. Lieb and F. Y. Wu, Phys. Rev. Lett. {\bf 20},
        1445 (1968).       
\bibitem[12]{Essler}
        Fabian H. L. Essler and Vladimir E. Korepin, 
        Phys. Rev. Lett. {\bf 72}, 908 (1994);
        {\it ibid.} Nucl. Phys. B {\bf 426},  (1994).             
\bibitem[13]{Faddeev}
        L. D. Faddeev and L. A. Takhtajan, 
        Phys. Lett. {\bf 85A}, 375 (1981).             
\bibitem[14]{Heilmann}
        O. J. Heilmann and E. H. Lieb, Ann. N. Y.
        Acad. Sci. {\bf 172}, 583 (1971).
\bibitem[15]{Lieb89}
        E. H. Lieb, Phys. Rev. Lett. {\bf 62},
        1201 (1989).       
\bibitem[16]{Yang89}
        C. N. Yang, Phys. Rev. Lett. {\bf 63}, 2144
        (1989).     
\bibitem[17]{Zhang}
        C. N. Yang and S. C. Zhang,
        Mod. Phys. Lett. B {\bf 4}, 759 (1990).
\bibitem[18]{Korepin}
        Fabian H. L. Essler, Vladimir E. Korepin, and
        Kareljan Schoutens, Phys. Rev. Lett. {\bf 67},
        3848 (1991); Nucl. Phys. B {\bf 372}, 559 (1992).             
\bibitem[19]{Ostlund}
        Stellan \"Ostlund, Phys. Rev. Lett. {\bf 69},
        1695 (1992).                                          
\bibitem[20]{Carmelo95}
        J. M. P. Carmelo and N. M. R. Peres, Phys. Rev. B
        {\bf 51}, 7481 (1995).
\bibitem[21]{Carmelo90}
        J. Carmelo and A. A. Ovchinnikov, Carg\`ese Lecture
        1990 (unpublished); J. Phys.: Condens. 
        Matter {\bf 3}, 757 (1991).
\bibitem[22]{Carmelo91}
        J. Carmelo, P. Horsch, P.-A. Bares, and A. A. Ovchinnikov, 
        Phys. Rev. B {\bf 44}, 9967 (1991).
\bibitem[23]{Carmelo92}
        J. M. P. Carmelo, P. Horsch, and A. A. Ovchinnikov, 
        Phys. Rev. B {\bf 45}, 7899 (1992).    
\bibitem[24]{Carmelo92b}
        J. M. P. Carmelo and P. Horsch,
        Phys. Rev. Lett. {\bf 68}, 871 (1992);
        J. M. P. Carmelo, P. Horsch, and A. A. Ovchinnikov, 
        Phys. Rev. B {\bf 46}, 14 728 (1992).    
\bibitem[25]{Carmelo93}        
        J. M. P. Carmelo, P. Horsch, D. K. Campbell, and
        A. H. Castro Neto, Phys. Rev. B {\bf 48}, 4200 (1993).
\bibitem[26]{Carmelo93b}
        J. M. P. Carmelo and A. H. Castro Neto,
        Phys. Rev. Lett. {\bf 70}, 1904 (1993).
\bibitem[27]{Carmelo94}
        J. M. P. Carmelo, A. H. Castro Neto, and 
        D. K. Campbell, Phys. Rev. B {\bf 50},
        3667 (1994). 
\bibitem[28]{Carmelo94b}
        J. M. P. Carmelo, A. H. Castro Neto, and 
        D. K. Campbell, Phys. Rev. B {\bf 50},
        3683 (1994).
\bibitem[29]{Haldane91}
        F. D. M. Haldane, Phys. Rev. Lett. {\bf 66}, 1529 (1991).
\bibitem[30]{Shastry}
        E. R. Mucciolo, B. Shastry, B. D. Simons, and
        B. L. Altshuler, Phys. Rev. B {\bf 49}, 15 197 (1994).
\bibitem[31]{Carmelo95b}
        J. M. P. Carmelo, A. H. Castro Neto, and N. M. R. Peres, 
        preprint (1995); {\it ibid.} unpublished.
\bibitem[32]{Carmelo95c}
        J. M. P. Carmelo, N. M. R. Peres, and D. K.
        Campbell, preprint (1995).
\bibitem[33]{Anto}
        A. H. Castro Neto, H. Q. Lin, Y. -H.
        Chen, and J. M. P. Carmelo, Phys. Rev. B {\bf 50}
        14 032 (1994).
\bibitem[34]{Frahm}
        Holger Frahm and V. E. Korepin, Phys. Rev. B {\bf 42},
        10 553 (1990); {\it ibid.} {\bf 43}, 5653 (1991).             
\bibitem[35]{Carmelo96}
        J. M. P. Carmelo and N. M. R. Peres,
        to appear in 1996.
\end{references}
\end{document}